\documentclass[aps,%
twocolumn,%
floatfix,%
pra,%
amsfonts,%
groupedaddress,%
superscriptaddress]{revtex4-2}%
\usepackage[utf8]{inputenc}
\usepackage[T1]{fontenc}
\usepackage{mathrsfs}
\usepackage{mathtools}
\usepackage{amsmath}
\usepackage{amsfonts}
\usepackage{amssymb}
\usepackage[titletoc,title]{appendix}
\usepackage{algorithm,algorithmic}

\newenvironment{algorithmstage}[1]{%
  \renewcommand{\algorithmicbody}{\textbf{#1}}\BODY
}{\ENDBODY}
\renewenvironment{algorithm}[1][t]{%
  \begin{figure}[#1]%
  \captionsetup{type=algorithm,name=Protocol,labelformat=boldprotocol}%
  \centering%
  \begingroup%
  \small%
  \setlength{\abovecaptionskip}{1pt}%
  \setlength{\belowcaptionskip}{1pt}%
  \let\algorithmoldcaption\caption%
  \renewcommand{\caption}[1]{%
    \algorithmoldcaption{##1}%
    \vspace{2pt}\hrule height 0.45pt\vspace{4pt}%
  }%
  \begin{minipage}{0.98\columnwidth}%
  \hrule height 0.65pt\vspace{2pt}%
}{%
  \vspace{2pt}\hrule height 0.65pt%
  \end{minipage}%
  \endgroup%
  \end{figure}%
}
\usepackage{bbm}
\usepackage{bm}
\usepackage{amsthm}
\usepackage{needspace}
\usepackage{makecell}
\usepackage{array}
\usepackage[dvipsnames,table,xcdraw]{xcolor}
\definecolor{beamer@blendedblue}{rgb}{0.2,0.2,0.7}
\usepackage[osf,sc]{mathpazo}
\usepackage{caption}
\DeclareCaptionLabelFormat{boldprotocol}{\textbf{#1} #2}
\usepackage{subfig}
\usepackage{graphicx}
\usepackage{overpic}
\usepackage{booktabs}
\usepackage{multirow}
\usepackage{pgfplots}
\pgfplotsset{compat=1.18}
\usepgfplotslibrary{fillbetween}
\usetikzlibrary{arrows.meta,calc,positioning,fit,shapes.geometric}
\newcolumntype{C}[1]{>{\centering\arraybackslash}p{#1}}
\usepackage[colorlinks=true,%
linkcolor=beamer@blendedblue,%
bookmarks=true,%
breaklinks=true,%
filecolor=beamer@blendedblue,%
anchorcolor=yellow,%
citecolor=beamer@blendedblue,%
urlcolor=beamer@blendedblue,%
pdfauthor={Kun Wang},%
pdfsubject={Two-stage verification of pure continuous-variable quantum states},%
CJKbookmarks=true]{hyperref}
\usepackage{float}
\usepackage{pifont}
\newcommand\yes{\ding{52}}
\newcommand\no{\ding{56}}

\newtheorem{theorem}{Theorem}
\newtheorem{lemma}{Lemma}

\newtheorem{corollary}[theorem]{Corollary}

\mathchardef\ordinarycolon\mathcode`\:
\mathcode`\:=\string"8000
\def\vcentcolon{\mathrel{\mathop\ordinarycolon}}
\begingroup \catcode`\:=\active
  \lowercase{\endgroup
  \let :\vcentcolon
  }

\DeclareFontFamily{U}{mathx}{\hyphenchar\font45}
\DeclareFontShape{U}{mathx}{m}{n}{<-> mathx10}{}
\DeclareSymbolFont{mathx}{U}{mathx}{m}{n}
\DeclareMathAccent{\widebar}{0}{mathx}{"73}

\newcommand{\ket}[1]{\vert{#1}\rangle}
\newcommand{\bra}[1]{\langle{#1}\vert}

\newcommand{\braket}[2]{\langle #1\vert #2\rangle}
\newcommand\proj[1]{\vert{#1}\rangle\!\langle{#1}\vert}

\DeclareMathOperator{\tr}{Tr}
\newcommand{\1}{\mathbbm{1}}

\newcommand*{\cH}{\mathcal{H}}
\newcommand*{\cI}{\mathcal{I}}

\newcommand*{\cO}{\mathcal{O}}

\definecolor{googleblue}{HTML}{4285F4}
\definecolor{googlered}{HTML}{DB4437}
\definecolor{googleyellow}{HTML}{F4B400}
\definecolor{googlegreen}{HTML}{0F9D58}

\colorlet{cvblue}{googleblue}
\definecolor{cvorange}{HTML}{E37400}
\definecolor{cvlightblue}{HTML}{D2E3FC}
\definecolor{cvlightyellow}{HTML}{FEEFC3}
\definecolor{cvlightgreen}{HTML}{CEEAD6}
\definecolor{cvlightred}{HTML}{FAD2CF}
\definecolor{cvlightgrey}{HTML}{F1F3F4}
\definecolor{cvgrey}{HTML}{9AA0A6}
\definecolor{cvblack}{HTML}{202124}
\tikzset{
  cvscheme/.style={x=1cm,y=1cm,every node/.style={font=\small,text=cvblack}},
  cvcard/.style={rounded corners=3pt,draw=cvgrey,line width=0.8pt,
    align=center,inner sep=6pt},
  cvschematiccard/.style={text width=2.95cm,inner xsep=4pt,minimum height=1.2cm},
  cvsource/.style={cvcard,cvschematiccard,draw=cvorange,fill=cvlightyellow},
  cvtest/.style={cvcard,draw=cvblue,fill=cvlightblue},
  cvdecision/.style={diamond,aspect=1.8,draw=cvgrey,fill=cvlightgrey,
    line width=0.8pt,align=center,inner sep=2pt},
  cvverdict/.style={cvcard,font=\small\bfseries,
    minimum width=1.45cm,minimum height=0.65cm},
  cvgood/.style={cvverdict,fill=cvlightred},
  cvbad/.style={cvverdict,fill=cvlightgreen},
  cvflow/.style={-{Latex[length=2.2mm]},line width=0.9pt,draw=cvblack},
  cvquantum/.style={cvflow,line width=1.1pt},
  cvclassical/.style={cvflow,line cap=round,dash pattern=on 0pt off 2.6pt}
}

\newcommand{\TMSV}{\mathrm{TMSV}}

\begin{document}

\newcommand{\thetitle}{Two-Stage Verification of Pure Continuous-Variable Quantum States}
\hypersetup{pdftitle={\thetitle},pdfsubject={Two-stage verification of pure continuous-variable quantum states}}
\title{\thetitle}
\author{Kun Wang}
\email{nju.wangkun@gmail.com}
\affiliation{College of Computer Science and Technology,
National University of Defense Technology, Changsha 410073, China}
\date{\today}

\begin{abstract}
As continuous-variable (CV) quantum technologies advance, experiments must establish whether the multimode states they prepare meet the fidelity requirements of the applications.
To perform this fidelity check, we develop a two-stage protocol that uses local operations and classical communication (LOCC) to verify any known multimode pure target state.
Our key observation is that a common cutoff can bound all the target's photon counts with arbitrarily high probability.
In Stage~1, the Fock tail test estimates how often a source copy yields a count above this cutoff in any mode.
The same cutoff defines the subspace for Stage~2, where discrete-variable (DV) state verification tests fresh copies for disagreement with the projected target.
We establish verification guarantees without assuming an energy bound on the source.
At fixed confidence and mode number, the sample complexity bound scales inversely with the infidelity threshold.
We demonstrate the protocol's versatility by applying it to diverse families of Gaussian and non-Gaussian states.
\end{abstract}
\maketitle

\section{Introduction}\label{sec:introduction}

Continuous-variable (CV) quantum information~\cite{Weedbrook2012Gaussian} 
is advancing rapidly across quantum computing~\cite{aghaee2025modular}, 
quantum communication~\cite{Furusawa1998Teleportation,Ren2026DenseCoding}, and 
quantum metrology~\cite{Guo2020DistributedSensing}. 
These applications depend on multimode entangled states of high quality that 
enable quantum teleportation~\cite{Furusawa1998Teleportation}, 
measurement-based quantum computing~\cite{Menicucci2006Cluster,Zhang2006Cluster}, 
secure communication~\cite{Xu2020Secure}, and 
entanglement-enhanced sensing~\cite{Guo2020DistributedSensing}. 
Experiments have progressed from four-mode Greenberger--Horne--Zeilinger and cluster states~\cite{Su2007Quadripartite} and eight-mode clusters~\cite{Su2012EightPartite} to time-multiplexed clusters comprising ten thousand 
and one million optical modes~\cite{Yokoyama2013Cluster,Yoshikawa2016Million}, followed by two-dimensional~\cite{Asavanant2019Cluster,Larsen2019Cluster} and three-dimensional~\cite{Roh2025Cluster} cluster states. Integrated photonics now supports multipartite entanglement in microcombs~\cite{jia2025microcomb} and monolithic generation, manipulation, and measurement of cluster states~\cite{jia2026monolithic}. Optical cat state resources have enabled preparation of logical states in propagating light~\cite{konno2024logical}, while integrated sources generate Gottesman--Kitaev--Preskill (GKP) qubits~\cite{larsen2025gkp}. Modular processors~\cite{aghaee2025modular} and nonlinear optical operations~\cite{yu2026nonlinearity} further expand the tools for bosonic quantum computing.

As these platforms grow in scale and complexity, their usefulness depends on maintaining the quality of the prepared resources throughout generation, transmission, and processing. Imperfect preparation and loss can reduce computational accuracy, compromise communication security, and limit the sensitivity of quantum measurements. Reliable characterization is therefore essential for calibration, optimization, and practical deployment~\cite{Eisert2020Certification,KlieschRoth2021Theory,Yu2022Statistical}. These advances make it increasingly urgent to verify whether experimentally prepared multimode entangled CV states meet 
the quality requirements of their intended applications. 

\emph{Quantum state verification}~\cite{Pallister2018Optimal} uses feasible measurements to test
whether an experimentally prepared state is sufficiently close to a known target state.
Discrete-variable (DV) verification has established theoretical frameworks and experimental techniques 
using local operations and classical communication (LOCC)~\cite{Pallister2018Optimal,HayashiMorimae2015Blind,ZhuHayashi2019Efficient,Liu2021Nondemolition,MiguelRamiro2022Collective,Yu2022Statistical,WangHayashi2019TwoQubit,Jiang2020Standardization,Chen2025Memory,Zheng2024Subspaces}.
In contrast, CV verification remains comparatively underexplored, possibly due to the technical difficulty of obtaining rigorous fidelity guarantees from finite experimental data in an infinite-dimensional Hilbert space.
Full tomography reconstructs more information than verification requires~\cite{LvovskyRaymer2009Tomography}, at prohibitive cost for generic states.
Mele \emph{et al.} estimate $\sim3000$ years to reconstruct a 10-mode pure state with trace distance error $0.1$, assuming a preparation and processing time of $1\,\mathrm{ns}$ per copy and an average oscillator energy bound of $E=1$ per mode in their units~\cite{Mele2025Learning}.
Fidelity estimation and witnesses avoid full reconstruction~\cite{daSilva2011Characterization,Aolita2015Certification,Upreti2024Framework,Chabaud2021CertificationNG}. Standard concentration bounds for additive fidelity estimation give sufficient copy budgets proportional to the inverse square of the precision.
Dedicated protocols exploit target structure to certify adversarial CV resources~\cite{Takeuchi2019Serfling,Wu2021EfficientCV}, with graph state tests accommodating finite squeezing and measurement resolution~\cite{Descamps2025Noisy}.
Reductions address correlated sources~\cite{Fawzi2024NonIID}, while related methods cover CV learning~\cite{Wu2024Learning}, similarity testing~\cite{Wu2023Similarity}, bosonic code certification~\cite{Wu2022BosonicCodes}, and verifiable blind computation~\cite{Liu2019Blind}.
For entangled coherent states, Liu \emph{et al.} give a protocol based on 
photon counting whose sample complexity scales inversely with the infidelity threshold~\cite{Liu2021CVQSV}.
Extending such efficient verification to broader CV targets remains challenging.
Reliable fidelity guarantees must account for preparation errors at arbitrarily high photon numbers without assuming a bound on the unknown source's energy.

\begin{figure*}[t]
\centering
\includegraphics[width=0.98\textwidth]{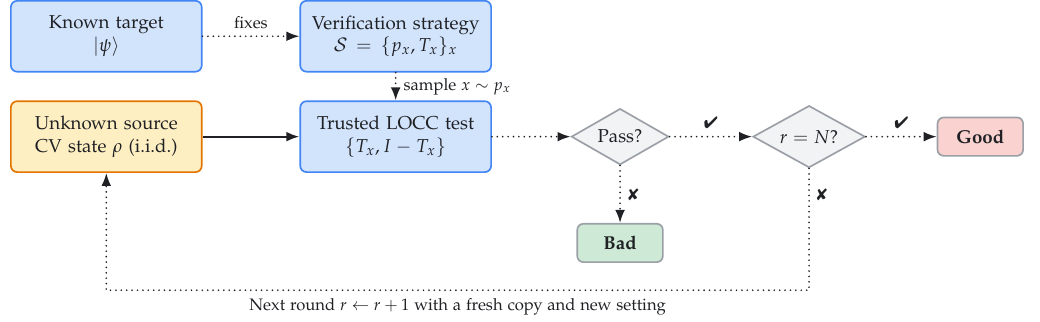}
\caption{\raggedright\textbf{Schematic view of continuous-variable quantum state verification.}
For each trusted binary LOCC test with setting $x$, let $p_x$ be its sampling probability and $T_x$ its pass effect, with $I$ the identity operator on all modes. The known target $\ket{\psi}$ determines the strategy $\mathcal{S}=\{p_x,T_x\}_x$.
The unknown source supplies independent, identically distributed copies of the $m$-mode state $\rho$. In round $r$, the parties share one setting $x$ drawn from $p_x$, independently of the draws in other rounds. A failure yields \textbf{Bad}, and passing all $N$ prescribed rounds yields \textbf{Good}.
Perfect completeness, $T_x\ket{\psi}=\ket{\psi}$ for every sampled setting, ensures that the exact target always receives \textbf{Good}.
Solid arrows carry quantum states, and dotted arrows carry classical information. The symbols \yes\ and \no\ denote true and false outcomes.}
\label{fig:cv-qsv}
\end{figure*}

\textbf{Contributions.}
We develop a two-stage LOCC protocol to verify an \emph{arbitrary} known multimode pure target.
Our key observation is that the target's photon statistics let us choose a common cutoff so that the probability of any count exceeding it is as small as desired.
Stage~1 uses this cutoff to estimate the fraction of source copies with a count above it in any mode.
The same cutoff defines the finite Fock subspace and thereby the \emph{projected target}, the target's normalized component in that subspace.
Stage~2 uses standard DV state verification~\cite{Yu2022Statistical} to test fresh copies for agreement with this state.
Together, the two stages give a fidelity guarantee for the full state without an energy bound on the source.

\textbf{Organization.} Sec.~\ref{sec:model} formulates the verification task and its CV obstruction. Sec.~\ref{sec:protocol} then presents the two-stage verification protocol together with its guarantees on completeness, soundness, and sample complexity. Secs.~\ref{sec:gaussian-applications} and~\ref{sec:non-gaussian-applications} apply the protocol to Gaussian and non-Gaussian states, respectively.

\section{Verification Task and the CV Obstruction}\label{sec:model}

This section defines the verification task and explains the obstruction posed by perfect target acceptance in CV systems. We first formulate the one-sided task, then examine its CV obstructions and introduce controlled honest rejection as a workaround.

\subsection{The state verification task}\label{sec:state-verification-task}

Consider a quantum light source designed to prepare a known normalized pure state $\ket{\psi}$ of $m$ bosonic modes distributed among $m$ local parties. The verifier has a classical description of this target.
Each mode $j$ has continuous field quadratures and a Hilbert space $\cH_j$ with photon number (Fock) basis $\{\ket{n}_j:n=0,1,\ldots\}$, where $n$ counts the photons in that mode. Each party holds one mode, so $\cH=\bigotimes_{j=1}^{m}\cH_j$ is the joint space.
A representative example is the two-mode squeezed vacuum (TMSV) state, whose entanglement enables continuous-variable quantum teleportation~\cite{Furusawa1998Teleportation} and dense coding~\cite{Ren2026DenseCoding}.
We assume that the quantum source is \emph{independent and identically distributed} (i.i.d.).
Under this assumption, $N$ emissions have joint state $\rho^{\otimes N}$ for an unknown density operator $\rho$.
Unlike the tomography setting in Ref.~\cite{Mele2025Learning}, 
we do not impose an energy constraint on the actually prepared state $\rho$. 

In functions of a state, a ket label denotes its density operator. For instance, $\psi=\proj{\psi}$.
The fidelity $F(\rho,\psi)$ is the probability that $\rho$ passes the ideal projection onto $\ket{\psi}$, and its infidelity is $1-F(\rho,\psi)$. The \emph{infidelity threshold} $0<\varepsilon<1$ is the smallest fidelity deficit that the verifier is required to detect.
Using finitely many copies, the verifier must distinguish the following two cases:
\begin{itemize}
\item \textbf{Good}: the source prepares the exact target, $\rho=\psi$.
\item \textbf{Bad}: the source prepares an $\varepsilon$-bad state, satisfying
\begin{align}
F\left(\rho,\psi\right):=\bra{\psi}\rho\ket{\psi}\leq 1-\varepsilon.
\label{eq:hypotheses}
\end{align}
\end{itemize}

In its original form, quantum state verification requires \emph{perfect completeness}~\cite{Pallister2018Optimal}. For any event $A$, ${\Pr}_\rho(A)$ denotes the probability of $A$ when the verifier tests independent copies of $\rho$, accounting for both its random measurement choices and the measurement outcomes. In particular, ${\Pr}_\rho(\mathrm{Good})$ is the probability that the verifier accepts. For an honest source preparing the exact target $\rho=\psi$, perfect completeness requires ${\Pr}_\psi(\mathrm{Good})=1$.

We assume trusted local implementations of the prescribed measurements. Each party may act jointly on its signal mode and locally prepared ancillas. The parties exchange only classical messages, which are trusted and may guide subsequent local operations.

Let $\mathcal X$ be the set of measurement settings and $I$ the identity on $\cH$.
For setting $x\in\mathcal X$, let $T_x$ be its pass effect, with $0\leq T_x\leq I$. The setting implements the binary LOCC test $\{T_x,I-T_x\}$.
A randomized verification strategy $\mathcal{S}=\{p_x,T_x\}_{x\in\mathcal X}$ chooses these settings with probabilities $p_x\geq0$ and $\sum_xp_x=1$. For each fresh copy, the parties share one draw $x$ and perform the corresponding test. The draws are independent across rounds, and the source does not learn $x$ before emitting the copy. In the standard one-sided protocol, the parties declare \textbf{Bad} as soon as a test fails, and declare \textbf{Good} if all $N$ rounds pass. Fig.~\ref{fig:cv-qsv} illustrates this procedure.
Note that the perfect completeness condition requires every test to accept the target with certainty, 
so for every test $T_x$ with $p_x > 0$, it holds that
\begin{align}
\bra{\psi}T_x\ket{\psi}=1
\quad\Longleftrightarrow\quad T_x\ket{\psi}=\ket{\psi}.
\label{eq:perfect-completeness}
\end{align}

The corresponding verification operator is written as
\begin{align}
\Omega=\sum_xp_xT_x,\qquad 0\leq\Omega\leq I.
\label{eq:general-verifier}
\end{align}
With $\tr$ denoting the trace, the probability that $\rho$ passes in one round is $p_{\rm pass}(\Omega,\rho)=\tr[\Omega\rho]$. Perfect completeness is equivalently expressed as $\Omega\ket{\psi}=\ket{\psi}$.

Write $\lVert\phi\rVert=\sqrt{\braket{\phi}{\phi}}$ for the Hilbert space norm of a vector $\ket{\phi}$.
For a bounded operator $A$, its operator norm is $\lVert A\rVert_\infty=\sup_{\lVert\phi\rVert=1}\lVert A\ket{\phi}\rVert$.
In finite dimensions this is the spectral norm, equal to the largest singular value of $A$.
To quantify rejection of bad sources, define the \emph{spectral gap} of $\Omega$ by
\begin{align}
\nu\left(\Omega\right)
:=1-\sup_{\substack{\ket{\phi}\perp\ket{\psi}\\\left\lVert\phi\right\rVert=1}}\bra{\phi}\Omega\ket{\phi}
=1-\left\lVert\Omega-\psi\right\rVert_\infty.
\label{eq:general-gap}
\end{align}
The gap measures the separation between the target eigenvalue $1$ and the spectrum of $\Omega$ on $\ket{\psi}^{\perp}$. Every state in this subspace is rejected with probability at least $\nu(\Omega)$ per round, so a larger gap reduces the required number of copies at fixed infidelity and confidence.
Since $\Omega\leq I-\nu(\Omega)(I-\psi)$, the worst-case probability of passing in one round is
\begin{align}
P\left(\Omega,\varepsilon\right)
:=\sup_{F\left(\rho,\psi\right)\leq1-\varepsilon}\tr[\Omega\rho]
=1-\nu\left(\Omega\right)\varepsilon.
\label{eq:worst-pass}
\end{align}
To approach the bound, place weight $1-\varepsilon$ on the target and weight $\varepsilon$ on an orthogonal state with pass probability arbitrarily close to $1-\nu(\Omega)$. The supremum need not be attained in the infinite-dimensional CV space.

Under the i.i.d. assumption, an $\varepsilon$-bad quantum source passes all $N$ rounds with probability 
at most $P(\Omega,\varepsilon)^N$. For $\nu(\Omega)>0$ and confidence $1-\delta$ with $0<\delta<1$, we require the probability of accepting any bad source to be at most $\delta$. Hence
the sample complexity, with one source copy per round, is
\begin{align}
N\left(\Omega,\varepsilon,\delta\right)
= \left\lceil\frac{\ln\delta}{\ln\left[1-\nu\left(\Omega\right)\varepsilon\right]}\right\rceil
\approx \left\lceil\frac{1}{\nu\left(\Omega\right)\varepsilon}\ln\frac{1}{\delta}\right\rceil.
\label{eq:general-copies}
\end{align}
Here the approximation holds for $\nu(\Omega)\varepsilon\ll1$ at fixed $\delta$ and follows from $\lim_{x\to0}\ln(1+x)/x=1$. Designing an efficient one-sided strategy thus amounts to minimizing $P(\Omega,\varepsilon)$, 
or equivalently maximizing the gap, over available tests satisfying perfect completeness. 

\subsection{Perfect completeness renders CV verification hard}\label{sec:perfect-completeness}

Eq.~\eqref{eq:perfect-completeness} permits rejection only on events of zero target probability. For a finitely squeezed Gaussian target, the joint homodyne density is strictly positive. For any normalizable pure target, the heterodyne density is positive almost everywhere. A binary test based on either such readout must therefore accept almost every outcome and hence every source (Appx.~\ref{app:perfect-completeness}).

Photon-number-resolving detection (PNRD) can reveal outcomes forbidden to the target. For example, unequal local photon counts rule out a two-mode squeezed vacuum. Equal counts are also consistent with an incoherent mixture of matched number states. The parties can also flag a copy if any local count exceeds a cutoff $K$. For any finite $K$, even an ideal two-mode squeezed vacuum with nonzero squeezing can raise this flag. To use these counts for verification, we must allow for their expected frequency on the target.

\begin{figure*}[t]
\centering
\includegraphics[width=0.98\textwidth]{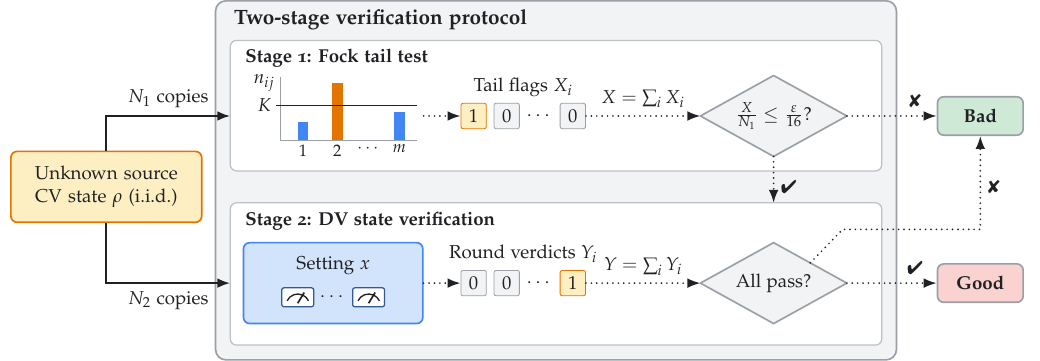}
\caption{\raggedright\textbf{Two-stage verification of a continuous-variable pure state.}
The gray box encloses both stages of Prot.~\ref{prot:two-stage}. They address the same verification task as the all-pass protocol in Fig.~\ref{fig:cv-qsv}.
In Stage~1, the bars show the photon counts $n_{ij}$ in mode $j$ of copy $i$. For each of the $N_1$ copies, the Fock tail test records $X_i=1$ if any count exceeds the common cutoff $K$ and $X_i=0$ otherwise. The protocol proceeds to Stage~2 if the flagged fraction satisfies $X/N_1\leq\varepsilon/16$.
In Stage~2 (DV state verification), the parties sample a setting $x$ for each of the $N_2$ fresh copies and perform the corresponding local measurements (meters). They combine their outcomes into a round verdict $Y_i$, with $0$ for a pass and $1$ for a rejection. The protocol returns \textbf{Good} if all Stage~2 rounds pass ($Y=0$) and \textbf{Bad} if either stage fails.}
\label{fig:two-stage}
\end{figure*}

\subsection{Relaxing perfect completeness}\label{sec:relaxing-completeness}

A perfectly complete test with a fixed positive spectral gap achieves the optimal $1/\varepsilon$ scaling of the sample complexity at fixed confidence~\cite{Pallister2018Optimal}.
Even the exact CV target, however, can produce photon counts above a finite cutoff. Relaxing perfect completeness lets us reject a source when such counts occur too often while bounding the probability of rejecting the target.

We keep the original i.i.d. source and the two hypotheses, $\rho=\psi$ and $F(\rho,\psi)\leq1-\varepsilon$. We relax Eq.~\eqref{eq:perfect-completeness} by allowing rejection of the exact target with probability at most $\delta_{\rm G}$ (Type~I error).
A separate budget $\delta_{\rm B}$ controls acceptance of the worst-case bad source (Type~II error). 
For $0\leq\delta_{\rm G}<1$ and $0<\delta_{\rm B}<1$, the final verdict must satisfy
\begin{align}
{\Pr}_\psi\left(\mathrm{Good}\right)\geq1-\delta_{\rm G},\;
\sup_{F\left(\rho,\psi\right)\leq1-\varepsilon}{\Pr}_\rho\left(\mathrm{Good}\right)\leq\delta_{\rm B}.
\label{eq:verification-errors}
\end{align}
For a specified number of source copies, we bound both error probabilities in Eq.~\eqref{eq:verification-errors}. At a given infidelity threshold, the \emph{sample complexity} is the number of source copies required to meet both error budgets in the worst case. A sufficient number of copies gives an upper bound on this quantity. At \emph{fixed confidence}, the budgets $\delta_{\rm G}$ and $\delta_{\rm B}$ are held fixed as the infidelity threshold $\varepsilon$ decreases. For a fixed target, we call the protocol \emph{sample-efficient} in this precision parameter if $\cO(1/\varepsilon)$ copies suffice, where $\cO(\cdot)$ denotes the big-$\cO$ notation.

Setting $\delta_{\rm G}=0$ recovers perfect completeness. Sources with $1-\varepsilon<F(\rho,\psi)<1$ remain outside the promise. Thinh, Dall'Arno, and Scarani~\cite{Thinh2020WorstCase} studied the corresponding single-copy problem for separable measurements. They minimized worst-case Type~II error subject to $\bra{\psi}\Omega\ket{\psi}\geq1-\delta_{\rm G}$ and found two-qubit regimes in which an optimal test does not commute with the target projector.

The verifier can preserve the inverse scaling by choosing measurements for the requested precision. For example, if the target failure probability is proportional to $\varepsilon$, then $\cO(1/\varepsilon)$ independent copies suffice to detect a fixed fractional increase in that rate at fixed confidence. Sec.~\ref{sec:protocol} gives conditions under which CV verification achieves this scaling while allowing occasional rejection of the exact target.

\section{Two-Stage Verification Protocol}\label{sec:protocol}

Here we describe a two-stage verification protocol for arbitrary known multimode pure CV states (Fig.~\ref{fig:two-stage}).
We choose a common cutoff $K$ from the target's photon statistics so that the probability of any target count exceeding it is as small as desired.
In Stage~1, the Fock tail test uses local PNRD to flag a source copy if any count exceeds $K$ (Sec.~\ref{sec:target-sector}).
Because even the exact target can produce flags, relaxing perfect completeness lets us reject a source when too many copies are flagged while controlling the probability of rejecting the target.
The same cutoff defines the finite Fock subspace in which Stage~2 uses standard DV state verification~\cite{Yu2022Statistical} to test fresh copies against the projected target (Sec.~\ref{sec:finite-sector}).
Sec.~\ref{sec:two-stage-protocol} combines the two tests and establishes the protocol's performance guarantees.

\subsection{Fock tail test}\label{sec:cutoff-selection}\label{sec:target-sector}

In one round of the \emph{Fock tail test} (FTT), each party measures the photon number of its mode by local PNRD. Let $n_{ij}$ be the count in mode $j$ on copy $i$. For a common nonnegative integer cutoff $K$, the parties report a single binary outcome
\begin{align}
X_i=\begin{cases}
0, & \max_j n_{ij}\leq K,\\
1, & \max_j n_{ij}>K.
\end{cases}
\label{eq:ftt-flag}
\end{align}
Thus $X_i=1$ flags the copy if at least one mode exceeds the cutoff. For mode $j$, let $I_j$ be the identity on $\cH_j$ and let $P_{K,j}=\sum_{n=0}^{K}\proj{n}_j$ be its cutoff projector. The joint projectors for $X_i=0$ and $X_i=1$, respectively, are
\begin{align}
P_{K}=\bigotimes_{j=1}^{m}P_{K,j},
\qquad P_K^{\perp}=I-P_{K}.
\label{eq:cutoff}
\end{align}
The Fock subspace $P_K\cH$ is spanned by the number states $\ket{n_1,\ldots,n_m}$ with $0\leq n_j\leq K$ for every mode and has dimension $(K+1)^m$. For any input state $\sigma$, the probability that copy $i$ is flagged ($X_i=1$) is
\begin{align}
q_K\left(\sigma\right):=\tr[P_K^{\perp}\sigma],\qquad
q_K(\psi)=\bra{\psi}P_K^{\perp}\ket{\psi}.
\label{eq:tails}
\end{align}

Now repeat the test on $N_1$ independent copies of the same source state $\rho$. The outcomes $X_i$ are independent Bernoulli variables with flag probability ${\Pr}_\rho(X_i=1)=q_K(\rho)$.
Write $\mathrm{Bin}(N,p)$ for the binomial distribution of the number of successes in $N$ independent trials, with success probability $p$ per trial.
The number $X$ of flagged copies thus satisfies
\begin{align}
X=\sum_{i=1}^{N_1}X_i\sim\mathrm{Bin}\left(N_1,q_K(\rho)\right).
\label{eq:ftt-count}
\end{align}
For an infidelity threshold $0<\varepsilon<1$, the decision rule based on the observed fraction of flagged copies is
\begin{align}
\text{Stage 1 outcome} &=
\begin{cases}
\text{pass}, & X/N_1\leq\varepsilon/16,\\
\text{fail}, & X/N_1>\varepsilon/16.
\end{cases}
\label{eq:ftt-decision}
\end{align}
If Stage~1 passes, the verifier proceeds to Stage~2. If it fails, the protocol stops and returns \textbf{Bad}.

We choose $K$ from the known target's photon statistics. For a joint count record $\boldsymbol n=(n_1,\ldots,n_m)$, write $\ket{\boldsymbol n}=\ket{n_1,\ldots,n_m}$ for its Fock state and $p_\psi(\boldsymbol n)=\lvert\braket{\boldsymbol n}{\psi}\rvert^2$ for its probability under the target. Then
\begin{align}
1-q_K(\psi)=\sum_{\boldsymbol n\in\left\{0,\ldots,K\right\}^{m}}p_\psi\left(\boldsymbol n\right)
          ={\Pr}_\psi\!\left(\max_j n_j\leq K\right).
\label{eq:cutoff-population}
\end{align}
As $K$ grows, the sum approaches one by normalization, even when the mode counts are correlated. We can therefore choose a finite common cutoff such that
\begin{align}
q_K(\psi)\leq\frac{\varepsilon}{32}.
\label{eq:design}
\end{align}
Let $\hat n_j$ be the photon number operator of mode $j$, with the subscript omitted for a single mode. If the local mean photon numbers $E_j=\bra{\psi}\hat n_j\ket{\psi}$ are finite and known, write $E_\psi=\sum_{j=1}^m E_j$ for the total mean photon number. The union bound and Markov's inequality then give a sufficient cutoff through
\begin{align}
q_K(\psi)\leq\sum_{j=1}^{m}{\Pr}_\psi\left(n_j>K\right)
\leq\frac{E_\psi}{K+1}.
\label{eq:target-markov}
\end{align}
Both choices fix $K$ from target data alone. Rare emissions with many photons can make the source's total mean photon number arbitrarily large even when its tail probability $q_K(\rho)$ is small (Appx.~\ref{app:mean-energy}).

Under the decision rule in Eq.~\eqref{eq:ftt-decision}, sampling fluctuations can cause a source with a small tail to fail Stage~1 or a source with an excessive tail to pass it. Lem.~\ref{lem:ftt} bounds both probabilities.

\begin{lemma}\label{lem:ftt}
Fix $0<\varepsilon<1$ and run the FTT with cutoff $K$ on $N_1$ i.i.d. copies of $\rho$. If $q_K(\rho)\leq\varepsilon/32$, the rejection probability satisfies
\begin{align}
{\Pr}_\rho\left(X/N_1>\varepsilon/16\right)\leq e^{-N_1\varepsilon/96}.
\label{eq:ftt-reject}
\end{align}
If $q_K(\rho)\geq\varepsilon/8$, the pass probability satisfies
\begin{align}
{\Pr}_\rho\left(X/N_1\leq\varepsilon/16\right)\leq e^{-N_1\varepsilon/64}.
\label{eq:ftt-pass}
\end{align}
\end{lemma}
By Eq.~\eqref{eq:design}, the exact target satisfies $q_K(\psi)\leq\varepsilon/32$. Eq.~\eqref{eq:ftt-reject} therefore bounds its probability of rejection by $\delta_{\rm G}$ when $N_1\geq(96/\varepsilon)\ln(1/\delta_{\rm G})$. 
For an unknown source with $q_K(\rho)\geq\varepsilon/8$, Eq.~\eqref{eq:ftt-pass} bounds the probability of proceeding to Stage~2 by $\delta_{\rm B}$ when $N_1\geq(64/\varepsilon)\ln(1/\delta_{\rm B})$. We choose $N_1$ to meet both bounds. Appx.~\ref{app:binomial} gives the proof of Lem.~\ref{lem:ftt}.

For Stage~2 we form the \emph{projected target} by projecting $\ket{\psi}$ onto $P_K\cH$ and normalizing. Since this outcome has probability $1-q_K(\psi)>0$, we obtain
\begin{align}
\ket{\psi_{K}}=\frac{P_{K}\ket{\psi}}{\sqrt{1-q_K(\psi)}}.
\label{eq:truncated-target}
\end{align}
Because the target $\ket{\psi}$ is known, the verifier can calculate $\ket{\psi_K}$ from Eq.~\eqref{eq:truncated-target} before testing the source. The tail flags give no information about which state a source prepares within $P_K\cH$.

\subsection{DV state verification}\label{sec:finite-sector}

If Stage~1 passes, Stage~2 uses a fresh batch of source copies. Its task within the finite Fock subspace $P_K\cH$ is a \emph{DV verification problem} for the projected target $\ket{\psi_K}$ in Eq.~\eqref{eq:truncated-target}.
We use the DV verification toolbox~\cite{Yu2022Statistical} to choose an LOCC strategy for this target, favoring the one with the largest certified spectral gap.
For the prescribed infidelity threshold $\varepsilon$ and false acceptance budget $\delta_{\rm B}$, a larger gap gives a smaller upper bound on the number of copies needed in Stage~2.
For each setting $x$, let $p_x$ be its sampling probability and $M_{x,K}$ its DV pass effect extended by zero outside $P_K\cH$. We denote the chosen strategy by $\mathcal{S}_K=\{p_x,M_{x,K}\}_x$ and use it throughout this stage.

In each round, the parties share a setting $x$ drawn with probability $p_x$ from $\mathcal{S}_K$. The fresh copy can have weight above the cutoff. The full-space test rejects only on records whose combined effect is the DV rejection effect $P_K-M_{x,K}$ supported inside the cutoff. The parties assign every other record to pass, including all records arising from a state supported entirely above the cutoff. The binary verdict is
\begin{align}
Y_i &=
\begin{cases}
0, & \text{pass, with effect } M_{x,K}+P_K^\perp,\\
1, & \text{reject, with effect } P_K-M_{x,K}.
\end{cases}
\label{eq:dv-round-outcome}
\end{align}
The pass effect follows by complementation, $I-(P_K-M_{x,K})=M_{x,K}+P_K^\perp$. The Stage~2 test on the full space uses the DV decision rule within $P_K\cH$ and accepts every state supported on the tail. A pass outcome alone leaves open whether the copy had weight above the cutoff.

The construction of the DV verification strategy $\mathcal{S}_K$ for $\ket{\psi_K}$~\cite{Yu2022Statistical} guarantees that its average pass effect $\Omega_K=\sum_x p_x M_{x,K}$ satisfies
\begin{subequations}\label{eq:finite-verifier}
\begin{align}
0&\leq\Omega_{K}\leq P_{K},\quad
\Omega_{K}\ket{\psi_{K}}=\ket{\psi_{K}},\\
\nu_{K}&=1-\left\lVert\Omega_{K}-\proj{\psi_{K}}\right\rVert_\infty>0.
\end{align}
\end{subequations}
Thus the projected target passes every sampled setting, and any state supported on its orthogonal complement within $P_K\cH$ fails with probability at least $\nu_K$. Averaging the full-space pass effect in Eq.~\eqref{eq:dv-round-outcome} over the settings gives the overall verification operator
$\Omega_{K}^+:=\Omega_{K}+P_K^{\perp}$ and the probability of a pass in one round,
\begin{align}
{\Pr}_\rho\left(Y_i=0\right)=\tr[\Omega_K^+\rho]=\tr[\Omega_K\rho]+q_K(\rho).\label{eq:lft-round}
\end{align}

Stage~2 essentially tests agreement with the projected target only within $P_K\cH$. In Eq.~\eqref{eq:lft-round}, $\tr[\Omega_K\rho]$ is the contribution to the pass probability from this subspace, and the entire tail weight $q_K(\rho)$ contributes to the same recorded pass. For the exact target, $\tr[\Omega_K\psi]=1-q_K(\psi)$, 
so the total pass probability is one. A source supported entirely above the cutoff also passes with certainty. Stage~2 therefore has zero verification gap on the full space, which is why Stage~1 tests whether the source places too much weight above the cutoff.

Now apply the DV test to $N_2$ independent copies of $\rho$, sampling the setting independently in each round. Let $Y=\sum_{i=1}^{N_2}Y_i$ count the failed rounds. The Stage~2 decision rule is
\begin{align}
\text{Stage 2 outcome} &=
\begin{cases}
\text{pass}, & Y=0,\\
\text{fail}, & Y\geq 1.
\end{cases}
\label{eq:dv-decision}
\end{align}
The protocol returns \textbf{Bad} at the first rejection and \textbf{Good} if all $N_2$ rounds pass. By independence, the pass probability in Eq.~\eqref{eq:dv-decision} is $[\tr[\Omega_K^+\rho]]^{N_2}$. Under the small-tail conditions of Lem.~\ref{lem:lft}, a source with $F(\rho,\psi)\leq1-\varepsilon$ has weight at least $\varepsilon/2$ in the part of $P_K\cH$ orthogonal to $\ket{\psi_K}$. The in-sector gap then gives a rejection probability of at least $\nu_K\varepsilon/2$ per round.

\begin{lemma}\label{lem:lft}
Fix $0<\varepsilon<1$ and a cutoff with $q_K(\psi)\leq\varepsilon/8$. On $N_2$ i.i.d. copies, apply the full-space test in Eq.~\eqref{eq:dv-round-outcome} exactly, using a DV test with gap $\nu_K$ as in Eq.~\eqref{eq:finite-verifier}. The exact target passes every round with certainty,
\begin{align}
{\Pr}_\psi\left(Y_i=0\right)={\Pr}_\psi\left(Y=0\right)=1.
\label{eq:lift-complete}
\end{align}
If $F(\rho,\psi)\leq1-\varepsilon$ and $q_K(\rho)<\varepsilon/8$, then
\begin{align}
{\Pr}_\rho\left(Y_i=0\right)&\leq 1-\nu_K\varepsilon/2,\label{eq:conditional-gap}\\
{\Pr}_\rho\left(Y=0\right)&\leq \left(1-\nu_K\varepsilon/2\right)^{N_2}
\leq e^{-N_2\nu_K\varepsilon/2}.\label{eq:lft-pass}
\end{align}
\end{lemma}
The cutoff in Eq.~\eqref{eq:design} satisfies the lemma's target condition. Taking $N_2\geq[2/(\nu_K\varepsilon)]\ln(1/\delta_{\rm B})$ therefore bounds the Stage~2 pass probability by $\delta_{\rm B}$ for a low-fidelity source with $q_K(\rho)<\varepsilon/8$. Appx.~\ref{app:tail-fidelity} gives the proof. For $q_K(\rho)\geq\varepsilon/8$, Lem.~\ref{lem:ftt} bounds the probability of passing Stage~1. Sec.~\ref{sec:two-stage-protocol} combines the two bounds to cover all low-fidelity sources.

\medskip\textbf{Optical implementation of the local DV measurements.}
The sample complexity results in Secs.~\ref{sec:gaussian-applications} and~\ref{sec:non-gaussian-applications} use the gaps of the specified ideal local DV tests. Appx.~\ref{app:dv-implementation} also gives an all-optical LOCC strategy using lossless local beam splitters, trusted coherent reference pulses, and ideal PNRD. Interference with a coherent reference makes photon detection sensitive to photon number coherence~\cite{Xu2020DetectorCoherence}. The optical strategy rejects on prescribed joint count records whose effects are supported within the cutoff. It gives a positive gap at each fixed cutoff, and its Stage~2 copy allocation uses that optical gap.

\subsection{Two-stage verification protocol}\label{sec:two-stage-protocol}\label{sec:guarantee}

We now combine the Fock tail test of Sec.~\ref{sec:target-sector} with DV state verification in Sec.~\ref{sec:finite-sector} to obtain a two-stage LOCC verification protocol for any known multimode pure target.
Prot.~\ref{prot:two-stage} gives the parameter choices and decision rules for the procedure shown schematically in Fig.~\ref{fig:two-stage}.
The two stages use independent batches of copies from the same source, and we assume that the parties can implement the full-space binary test in Eq.~\eqref{eq:dv-round-outcome} exactly by LOCC.
For inverse-infidelity scaling at fixed confidence, we choose a DV strategy whose spectral gap $\nu_K$ stays bounded away from zero as the cutoff grows. At fixed mode number, this is possible in the ideal local measurement model. For $m\geq2$, the adaptive strategy of Li and Zhu~\cite{Li2026Universal} guarantees $\nu_K\geq2^{1-m}$ for every projected pure target. For one mode, an ideal projection onto $\ket{\psi_K}$ gives $\nu_K=1$.

\begin{algorithm}[t]
\captionsetup{justification=raggedright,singlelinecheck=false}
\caption{Two-stage CV state verification}
\label{prot:two-stage}
\begin{tabular}{@{}l@{\hspace{0.35em}}l@{}}
\textbf{Input:} & \parbox[t]{0.82\linewidth}{\raggedright Known pure target $\ket{\psi}$ on $m$ bosonic modes (one per party), i.i.d. copies of $\rho$ on the same modes, infidelity threshold $\varepsilon$, and error budgets $\delta_{\rm G}$ (false rejection) and $\delta_{\rm B}$ (false acceptance), with $0<\varepsilon,\delta_{\rm G},\delta_{\rm B}<1$.}\\
\textbf{Output:} & \textbf{Good} (accept) or \textbf{Bad} (reject).
\end{tabular}
\vspace{-4pt}
\begin{algorithmic}[1]
\raggedright
\STATE Choose a positive integer $K$ with $q_K(\psi)\leq\varepsilon/32$, using the target tail or a certified upper bound on it.
\STATE Compute the projected target $\ket{\psi_K}=P_K\ket{\psi}/\sqrt{1-q_K(\psi)}$, a DV state in the finite Fock subspace $P_K\cH$.
\STATE Choose a DV LOCC verification protocol for $\ket{\psi_K}$ with settings $\{p_x,M_{x,K}\}_x$ satisfying Eq.~\eqref{eq:finite-verifier}.
\STATE Form $\Omega_K=\sum_x p_x M_{x,K}$ and compute its gap $\nu_K$ in Eq.~\eqref{eq:finite-verifier}.
\STATE Set $N_1\gets\lceil 1/\varepsilon\max\{96\ln(1/\delta_{\rm G}),64\ln(1/\delta_{\rm B})\}\rceil$.
\STATE Set $N_2\gets\lceil 2\ln(1/\delta_{\rm B})/(\nu_K\varepsilon)\rceil$. 
\begin{algorithmstage}{Stage 1. Fock tail test.}
    \STATE Initialize the running total for $X=\sum_{i=1}^{N_1}X_i$ to zero.
    \FOR{$i=1,\ldots,N_1$}
        \STATE On copy $i$, use local PNRD to record the photon count $n_{ij}$ in each mode $j=1,\ldots,m$.
        \STATE Set $X_i=1$ if $\max_j n_{ij}>K$ and $X_i=0$ otherwise. Update $X\gets X+X_i$.
        \STATE If $X/N_1>\varepsilon/16$, \textbf{return Bad}.
    \ENDFOR
\end{algorithmstage}
\begin{algorithmstage}{Stage 2. DV state verification.}
    \FOR{$i=1,\ldots,N_2$}
        \STATE On fresh copy $i$ of this batch, independently draw setting $x$ with probability $p_x$.
        \STATE Execute the local measurement and communication sequence specified by $x$, 
              including any outcome-dependent measurements.
              Reject on the prescribed records with total effect $P_K-M_{x,K}$ 
              nd pass on all other records, as in Eq.~\eqref{eq:dv-round-outcome}.
        \STATE Record $Y_i=1$ on rejection and $Y_i=0$ on a pass.
        \STATE If $Y_i=1$, \textbf{return Bad}.
    \ENDFOR
    \RETURN \textbf{Good} (all rounds passed, $Y=\sum_{i=1}^{N_2}Y_i=0$).
\end{algorithmstage}
\end{algorithmic}
\end{algorithm}

Thm.~\ref{thm:two-stage} gives the completeness, soundness, and sample complexity of Prot.~\ref{prot:two-stage}. By Lem.~\ref{lem:lft}, the exact target passes Stage~2 with certainty, so its rejection probability is determined by Stage~1. For a source with $F(\rho,\psi)\leq1-\varepsilon$, the tail probability determines which test controls acceptance. If $q_K(\rho)\geq\varepsilon/8$, Lem.~\ref{lem:ftt} bounds the probability of passing Stage~1. If $q_K(\rho)<\varepsilon/8$, Lem.~\ref{lem:lft} bounds the probability of passing Stage~2.

\begin{theorem}\label{thm:two-stage}
Prot.~\ref{prot:two-stage} accepts the exact target $\ket{\psi}$ with probability at least $1-\delta_{\rm G}$. For every source state $\rho$ with $F(\rho,\psi)\leq1-\varepsilon$, the probability of acceptance is at most $\delta_{\rm B}$. Using the gap $\nu_K$ of the chosen DV strategy, a sufficient copy budget as $\varepsilon\to0$ at fixed confidence is
\begin{align}
N=\cO\!\left(\frac{1}{\varepsilon}
\max\left\{\ln\left(1/\delta_{\rm G}\right),
\frac{\ln\left(1/\delta_{\rm B}\right)}{\nu_K}\right\}\right).
\label{eq:total-copies}
\end{align}
Thus at fixed confidence, $\cO(1/\varepsilon)$ source copies suffice if $\nu_K$ stays bounded away from zero as the cutoff grows.
\end{theorem}

Appx.~\ref{app:complete-protocol} gives the proof of Thm.~\ref{thm:two-stage}.
Several important remarks on Thm.~\ref{thm:two-stage} are in order:
\begin{enumerate}
\item \textbf{A nonzero target tail precludes perfect completeness in Prot.~\ref{prot:two-stage}.}
Stage~2 always accepts the exact target. If $q_K(\psi)>0$, however, there is a positive probability that all copies in a finite batch are flagged, causing Stage~1 to reject the target. At fixed $\varepsilon$ and $\delta_{\rm B}$, the protocol can attain arbitrarily small $\delta_{\rm G}>0$ by increasing the Stage~1 sample size logarithmically in $1/\delta_{\rm G}$. If $q_K(\psi)=0$, both stages accept the target with certainty.

\item \textbf{The Stage~1 sample size is independent of the target.}
The target's photon statistics determine the cutoff $K$ in Eq.~\eqref{eq:design}. Once that condition holds, $N_1$ depends only on $\varepsilon$, $\delta_{\rm G}$, and $\delta_{\rm B}$. A larger cutoff increases the dimension of the DV verification problem in Stage~2 while keeping the Stage~1 allocation fixed. These guarantees require no energy bound on the source.

\item \textbf{With equal error budgets, the DV gap controls the sample complexity bound.}
When $\delta_{\rm G}=\delta_{\rm B}=\delta$, the Stage~2 term attains the maximum in Eq.~\eqref{eq:total-copies} because $0<\nu_K\leq1$. The resulting bound is $\cO\!(\ln(1/\delta)/(\nu_K\varepsilon))$. With the exact gap and before rounding, the ratio of Stage~2 to Stage~1 copies is $1/(48\nu_K)$, so Stage~2 has the larger allocation precisely when $\nu_K<1/48$. The same calculation applies when a certified lower bound replaces the gap.

\item \textbf{Each stage can use the full false-acceptance budget.}
An $\varepsilon$-bad source with $q_K(\rho)\geq\varepsilon/8$ passes the Fock tail test with probability at most $\delta_{\rm B}$. Below this threshold, at least $\varepsilon/2$ of the source weight lies inside the cutoff and orthogonal to $\ket{\psi_K}$, so the probability of passing all DV rounds is at most $\delta_{\rm B}$. Since acceptance requires both stages to pass, either case gives the same overall bound $\delta_{\rm B}$.
\end{enumerate}

\section{Applications to Gaussian States}\label{sec:gaussian-applications}

We first apply Prot.~\ref{prot:two-stage} to arbitrary pure multimode Gaussian states. We then specialize the construction to two-mode states, whose canonical form gives the exact target photon tail and the gap of an explicit bipartite DV verification protocol.

Throughout Secs.~\ref{sec:gaussian-applications} and~\ref{sec:non-gaussian-applications}, we set $\delta_{\rm G}=\delta_{\rm B}=\delta$, with $0<\delta<1$.
For each target $\ket\psi$, the same $\delta$ bounds both the probability of rejecting the exact target and the probability of accepting any $\varepsilon$-bad source $\rho$ with $F(\rho,\psi)\leq1-\varepsilon$.
The common Stage~1 sample size is therefore
\begin{align}
N_1=\left\lceil\frac{96}{\varepsilon}\ln\frac{1}{\delta}\right\rceil.
\label{eq:examples-ftt-samples}
\end{align}
The Stage~2 allocation uses the same logarithmic dependence on $1/\delta$. When a certified lower bound on the Stage~2 gap is available, we use it in the allocation.
For unrounded stage allocations $a$ and $b$, the inequality $\lceil a\rceil+\lceil b\rceil\leq\lceil a+b\rceil+1$ gives the sufficient totals below.
We state the asymptotic sample complexity bounds as $\varepsilon\to0$ at fixed $\delta$.
If the DV verification gap $\nu_K$ stays bounded away from zero in this limit, $\cO(1/\varepsilon)$ source copies suffice.
Tab.~\ref{tab:family-sample-complexity} summarizes the Stage~2 spectral gaps and sample complexity bounds for both kinds of targets. 

\begin{table*}[t]
\centering
\small
\renewcommand{\arraystretch}{1.65}
\setlength{\tabcolsep}{3pt}
\begin{tabular*}{0.98\textwidth}{@{\extracolsep{\fill}}llccc@{}}
\toprule
\multirow{2}{*}{\textbf{Target class}} &
\multirow{2}{*}{\textbf{Pure target family}} &
\multicolumn{2}{c}{\textbf{Verification performance}} &
\multirow{2}{*}{\textbf{Result}} \\
& & Stage~2 spectral gap & \makecell{Sample complexity} & \\
\midrule
\multirow{2}{*}[-4pt]{Gaussian} & Arbitrary multimode Gaussian &
$\nu_K\geq 2^{1-m}$ & $\cO\!(2^m/\varepsilon)$ & Cor.~\ref{cor:multimode-gaussian-verification} \\
\addlinespace[2.5pt]
& Entangled two-mode Gaussian &
$\nu_K=\dfrac{\sum_{n=0}^K\tanh^{2n}r}{1+\sum_{n=0}^K\tanh^{2n}r}$ & $\cO\!(1/\varepsilon)$ & Cor.~\ref{cor:gaussian-verification} \\
\noalign{%
  \vskip 3pt
  \hbox{\tikz{\draw[cvgrey,line width=0.7pt,line cap=round,
    dash pattern=on 0pt off 2.6pt] (0,0) -- (0.98\textwidth,0);}}%
  \vskip 3pt
}
\multirow{7}{*}[-12pt]{Non-Gaussian} & TMSV with symmetric photon subtraction &
$\nu_K\geq\dfrac{S_K}{S_K+\max_{0\leq n\leq K}u_n}$ & $\cO\!(1/\varepsilon)$ & Cor.~\ref{cor:photon-subtraction} \\
\addlinespace[2.5pt]
& TMSV with symmetric photon addition &
$\nu_K\geq\dfrac{S_J}{S_J+\max_{0\leq n\leq J}u_n}$ & $\cO\!(1/\varepsilon)$ & Cor.~\ref{cor:photon-addition} \\
\addlinespace[2.5pt]
& Multimode entangled coherent state &
$\nu_K\geq 2^{1-m}$ & $\cO\!(2^m/\varepsilon)$ & Cor.~\ref{cor:ecs-verification} \\
\addlinespace[2.5pt]
& Balanced multimode NOON state &
$\nu_K\geq 1/m$ & $\cO\!(m/\varepsilon)$ & Cor.~\ref{cor:noon} \\
\addlinespace[2.5pt]
& Finite-energy GKP Bell pair &
$\nu_K\geq 1/2$ & $\cO\!(1/\varepsilon)$ & Cor.~\ref{cor:gkp-bell} \\
\addlinespace[2.5pt]
& Three-mode squeezed CV hypergraph state &
$\nu_K\geq 1/4$ & $\cO\!(1/\varepsilon)$ & Cor.~\ref{cor:cv-hypergraph} \\
\addlinespace[2.5pt]
& Finite-energy single-mode non-Gaussian state &
$\nu_K=1$ & $\cO\!(1/\varepsilon)$ & Cor.~\ref{cor:single-mode} \\
\bottomrule
\end{tabular*}
\caption{\raggedright\textbf{Sample complexity of two-stage verification for the target families in Secs.~\ref{sec:gaussian-applications} and~\ref{sec:non-gaussian-applications}.}
The target class refers to the original CV state before Fock projection. The in-sector gaps and sample complexity bounds refer to the specified ideal Stage~2 tests.
At the common error budget $\delta$, a sufficient total is $\lceil(96+2/\nu_K)\ln(1/\delta)/\varepsilon\rceil+1$ source copies. A listed lower bound on $\nu_K$ can be substituted in this formula. Here $m$ is the number of modes and $K$ is the photon cutoff.
TMSV denotes a two-mode squeezed vacuum with squeezing parameter $r$. Write $\ell$ for the number of photons subtracted or added per mode and $J=K-\ell$ for the shifted cutoff used for addition. Using binomial coefficients, define the Fock weights $u_n=\binom{n+\ell}{\ell}^{\!2}\tanh^{2n}r$ and their partial sums $S_j=\sum_{n=0}^j u_n$.}
\label{tab:family-sample-complexity}
\end{table*}

\subsection{Arbitrary pure multimode Gaussian states}\label{sec:multimode-Gaussian-states}

\paragraph{Target state.}
Let $\ket G$ be a known pure state on $m\geq2$ modes with a Gaussian Wigner function. Its first moments and covariance matrix specify the target~\cite{Weedbrook2012Gaussian}. This class includes squeezed entangled states for CV teleportation and multimode cluster states for measurement-based quantum computation~\cite{Furusawa1998Teleportation,Menicucci2006Cluster,Larsen2019Cluster}.

Finitely squeezed multimode graph states belong to this class. They are prepared from a product of squeezed vacua by Gaussian controlled-phase gates along the graph edges~\cite{Menicucci2011Graphical}. The graph weights and input squeezing determine the covariance matrix, and the first moments vanish. These data specify the local preprocessing and photon cutoff below, so the verification guarantees of Cor.~\ref{cor:multimode-gaussian-verification} apply directly.

\paragraph{Two-stage verification.}
Each party chooses a single-mode Gaussian unitary $L_j$ that brings its target marginal to thermal form~\cite{Weedbrook2012Gaussian}. Applying $L=\bigotimes_j L_j$ to every source copy before both stages preserves its fidelity with the correspondingly transformed target $\vert\widetilde G\rangle=L\ket G$.
For Stage~1, let $\bar n_j$ be the transformed marginal mean photon number in mode $j$ and set $\lambda_j=\bar n_j/(1+\bar n_j)$. The count distribution is
\begin{align}
{\Pr}_{\widetilde G}\left(n_j=n\right)=\left(1-\lambda_j\right)\lambda_j^n.
\label{eq:graph-marginal-photons}
\end{align}
The modes can be correlated. A union bound gives
\begin{align}
q_K(\widetilde G)\leq B_K:=\sum_j\lambda_j^{K+1}
\leq m\lambda_{\max}^{K+1},
\label{eq:graph-tail}
\end{align}
where $\lambda_{\max}=\max_j\lambda_j$.
For $0<\lambda_{\max}<1$, the upper bound in Eq.~\eqref{eq:graph-tail} gives the sufficient positive cutoff
\begin{align}
K=\max\left\{1,\left\lceil
\frac{\ln\left(32m/\varepsilon\right)}
{\ln\left(1/\lambda_{\max}\right)}
\right\rceil-1\right\}.
\label{eq:graph-cutoff}
\end{align}
This choice ensures $B_K\leq\varepsilon/32$. If $\lambda_{\max}=0$, each transformed marginal is vacuum and $K=1$ gives $B_K=0$.

For Stage~2, apply the adaptive LOCC verification protocol of Li and Zhu~\cite{Li2026Universal} to the projected $m$-partite pure target of local dimension $K+1$. The strategy has a cutoff-independent lower bound on its gap,
\begin{align}
\nu_K\geq2^{1-m}.
\label{eq:graph-general-gap}
\end{align}

The thermal tail bound sets a sufficient cutoff, and the DV gap bound depends only on the number of modes. Cor.~\ref{cor:multimode-gaussian-verification} gives the resulting sample complexity, with proof in Appx.~\ref{app:multimode-gaussian-proof}.

\begin{corollary}
\label{cor:multimode-gaussian-verification}
For any known pure Gaussian target $\ket G$ on $m\geq2$ modes, Prot.~\ref{prot:two-stage} with the local preprocessing above accepts $\ket G$ with probability at least $1-\delta$ and any $\varepsilon$-bad source with probability at most $\delta$. A sufficient number of source copies is
\begin{align}
N_{\rm Gauss}=\left\lceil\frac{96+2^m}{\varepsilon}\ln\left(1/\delta\right)\right\rceil+1.
\label{eq:multimode-gaussian-samples}
\end{align}
\end{corollary}

\subsection{Arbitrary pure two-mode Gaussian states}
\label{sec:two-mode-Gaussian-states}

For an entangled pure two-mode Gaussian target, local Gaussian preprocessing yields a TMSV with matched photon counts. The joint tail equals the tail of either mode, so we can choose the smallest admissible cutoff and hence the smallest local DV dimension in this frame. The projected target's Schmidt probabilities yield an explicit one-way LOCC test whose gap exceeds the general lower bound of $1/2$, 
tightening the Stage~2 sample complexity bound in Cor.~\ref{cor:multimode-gaussian-verification}.

\paragraph{Target state.}
Label the two modes $A$ and $B$. For finite squeezing $r\geq0$, set $\lambda=\tanh^2r$. With the squeezing phase set to zero, the TMSV has the Fock expansion
\begin{align}
\ket{\TMSV\left(r\right)}=\sqrt{1-\lambda}
\sum_{n=0}^{\infty}\lambda^{n/2}\ket{n,n}.
\label{eq:tmsv}
\end{align}
The TMSV provides the entanglement for CV teleportation and dense coding~\cite{Furusawa1998Teleportation,Ren2026DenseCoding}.
For squeezing phase $\phi$, the coefficient of $\ket{n,n}$ acquires a factor $e^{in\phi}$. A local phase rotation $e^{-i\phi\hat n_A}$ on mode $A$ therefore brings a target with known $\phi$ to the form in Eq.~\eqref{eq:tmsv}.

Any pure two-mode Gaussian target $\ket G$ admits a canonical form with single-mode Gaussian unitaries $U_A$ and $U_B$ and some TMSV squeezing parameter $r\geq0$~\cite[Thm.~1]{Hanamura2025Beyond},
\begin{align}
\ket G=\left(U_A\otimes U_B\right)\ket{\TMSV\left(r\right)}.
\label{eq:gaussian-canonical}
\end{align}
Here $r$ is the squeezing parameter associated with $\ket G$.
The product target at $r=0$ is straightforward to verify. Apply the local inverses $U_A^\dagger$ and $U_B^\dagger$ and count photons in both modes. A copy passes if both counts are zero, which occurs with probability $F(\rho,G)$. We therefore focus on the entangled case $r>0$ in the two-stage construction below.

\paragraph{Two-stage verification.}
To verify $\ket G$, set $U=U_A\otimes U_B$ and apply the local inverse $U^\dagger$ to every source copy before either stage. The transformed source state $\widetilde\rho=U^\dagger\rho U$ satisfies
\begin{align}
F(\rho,G)=F(\widetilde\rho,\TMSV(r)).
\label{eq:gaussian-fidelity-reduction}
\end{align}
The transformed source remains i.i.d. and the tests remain LOCC. In the original frame, the cutoff projector and the full-space pass effect are
\begin{align}
P_{K,G}=UP_KU^\dagger,\qquad
\Omega_{K,G}^+=U(\Omega_K+P_K^{\perp})U^\dagger.
\label{eq:gaussian-lift}
\end{align}
For Stage~1, count photons after the inverse transformations. The target gives identical counts in both modes, so the joint tail is exactly
\begin{align}
q_K(\TMSV(r))=\lambda^{K+1}.
\label{eq:tmsv-tail}
\end{align}
For $0<\lambda<1$, the least positive cutoff satisfying the Stage~1 tail bound is
\begin{align}
K=\max\left\{1,\left\lceil\frac{\ln\left(32/\varepsilon\right)}{\ln\left(1/\lambda\right)}\right\rceil-1\right\}.
\label{eq:tmsv-cutoff}
\end{align}
At fixed squeezing, the cutoff grows logarithmically in $1/\varepsilon$. At large $r$ it grows as $\cO(e^{2r}\ln(1/\varepsilon))$.

For Stage~2, the normalized projected target has known Schmidt probabilities
\begin{align}
\ket{\psi_K}=\sum_{n=0}^K\sqrt{p_n}\ket{n,n},\qquad
p_n=\frac{\left(1-\lambda\right)\lambda^n}{1-\lambda^{K+1}}.
\label{eq:tmsv-truncated}
\end{align}
We adapt the optimal one-way protocol in Eqs.~(23) and~(24) of Yu, Shang, and G\"uhne~\cite{Yu2019Optimal} to the projected target. Its Schmidt basis test checks for matching photon counts. In the Fourier test, $A$ measures in a Fourier basis with randomized phases and $B$ tests the target's conditional state. Appx.~\ref{app:two-mode-Gaussian-states} specifies both tests.
Define $P_{\rm diag}=\sum_{n=0}^K\proj{n,n}$ and $w_K=p_0/(1+p_0)$. Let $\Omega_K^{\rm F}$ be the pass effect averaged over the Fourier settings. The average pass effect of the protocol is
\begin{subequations}\label{eq:tmsv-finite-verifier}
\begin{align}
\Omega_K&=w_KP_{\rm diag}+\left(1-w_K\right)\Omega_K^{\rm F},\\
\Omega_K^{\rm F}&=\proj{\psi_K}+\sum_{\substack{i,j=0\\i\ne j}}^K p_j\proj{i,j}.
\end{align}
\end{subequations}
At the same mixing weight, the one-way strategy in Eq.~(14) of Li, Han, and Zhu~\cite{Li2019Efficient} yields the same average pass effect. On the diagonal subspace orthogonal to $\ket{\psi_K}$, $\Omega_K$ has eigenvalue $w_K$. On $\ket{i,j}$ with $i\ne j$, its eigenvalue is $(1-w_K)p_j\leq w_K$. Hence, for $K\geq1$,
\begin{align}
\nu_K=\frac{1}{1+p_0}\geq\frac12.
\label{eq:tmsv-uniform-gap}
\end{align}

The exact TMSV tail sets the cutoff, and the largest projected Schmidt probability $p_0$ controls the Stage~2 allocation. Cor.~\ref{cor:gaussian-verification} gives the sample complexity for the original Gaussian target. Appx.~\ref{app:gaussian-proof} proves the result.

\begin{corollary}
\label{cor:gaussian-verification}
For any pure two-mode Gaussian target $\ket G$ with squeezing parameter $r>0$, Prot.~\ref{prot:two-stage} accepts $\ket G$ with probability at least $1-\delta$ and any $\varepsilon$-bad source with probability at most $\delta$. With the cutoff $K$ from Eq.~\eqref{eq:tmsv-cutoff}, a sufficient number of source copies is
\begin{align}
N_{\rm TMSV}=\left\lceil
\left(98+\frac{2\left(1-\tanh^2 r\right)}{1-\tanh^{2K+2}r}\right)
\frac{\ln(1/\delta)}{\varepsilon}
\right\rceil+1.
\label{eq:gaussian-samples}
\end{align}
\end{corollary}

\begin{figure}[t]
\centering
\includegraphics[width=\columnwidth]{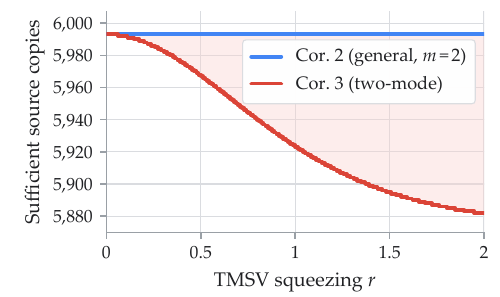}
\caption{\raggedright\textbf{Sample complexity bounds for entangled two-mode Gaussian states.} The curves show $N_{\rm Gauss}$ from Cor.~\ref{cor:multimode-gaussian-verification} at $m=2$ and $N_{\rm TMSV}$ from Cor.~\ref{cor:gaussian-verification} as functions of $r>0$, with $\varepsilon=\delta=0.05$. The two-mode bound uses the cutoff in Eq.~\eqref{eq:tmsv-cutoff}.}
\label{fig:two-mode-gaussian-comparison}
\end{figure}

\paragraph{Comparison with Cor.~\ref{cor:multimode-gaussian-verification}.}
The two-mode construction in Cor.~\ref{cor:gaussian-verification} can reduce the cutoff and tighten the Stage~2 sample complexity bound relative to the general construction in Cor.~\ref{cor:multimode-gaussian-verification}.
For example, at $\lambda=1/2$ and $\varepsilon=0.05$, the cutoff decreases from $10$ to $9$.
For $N_{\rm Gauss}$ at $m=2$, the coefficients of $\ln(1/\delta)/\varepsilon$ before rounding are $4$ for Stage~2 and $100$ in total.
The corresponding coefficients for $N_{\rm TMSV}$ are $2(1+p_0)$ and $98+2p_0$, respectively.
The fractional reduction is $(1-p_0)/2$ for Stage~2 and $(1-p_0)/50$ overall.
At fixed $\varepsilon$, these reductions approach $50\%$ and $2\%$ as $r$ increases.
At $r=2$ and $\varepsilon=\delta=0.05$, the sufficient total drops from $N_{\rm Gauss}=5\,993$ to $N_{\rm TMSV}=5\,882$ copies (Fig.~\ref{fig:two-mode-gaussian-comparison}).

\begin{figure}[t]
\centering
\includegraphics[width=\columnwidth]{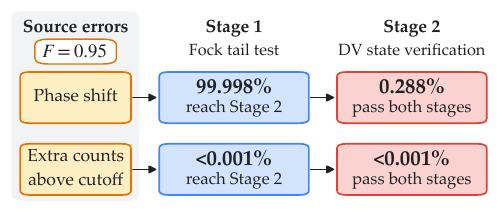}
\caption{\raggedright\textbf{Pass rates at equal fidelity.}
Both sources have fidelity $F=0.95$ to the TMSV target with $\tanh^2r=1/2$. For each source, the percentages give the probabilities of reaching Stage~2 and of passing both tests in one application of Prot.~\ref{prot:two-stage}. Stage~2 accounts for nearly all rejections of the phase-shifted source. Stage~1 accounts for all rejections of the source with extra counts above the cutoff.}
\label{fig:tmsv-error-acceptance}
\end{figure}

\paragraph{Acceptance under preparation errors.}
To complement the sample complexity comparison, we fix the copy allocation and examine 
how the two tests in the two-stage protocol contribute to rejection for different preparation errors at the same fidelity. In Fig.~\ref{fig:tmsv-error-acceptance}, both sources have fidelity $F=1-\varepsilon=0.95$ with respect to a common TMSV target~\eqref{eq:tmsv}. 
With $\varepsilon=\delta=0.05$, Prot.~\ref{prot:two-stage} uses the cutoff $K=9$ and allocates $N_1=5\,752$ and $N_2=180$ copies to the two tests. A run proceeds to Stage~2 if the Fock tail test flags at most $17$ copies, where a flag indicates a count above $K$ in either mode.

The first source prepares the target with a phase shift of about $0.16$ radians on mode $A$. The shift preserves the photon count distribution, so a full Stage~1 batch has about six flags on average and nearly all runs advance to Stage~2. Within the cutoff, the relative phases differ from those of the projected target. The DV test detects this difference, leaving about three in a thousand runs accepted overall.
The second source emits the target on $95\%$ of copies and the orthogonal Fock state $\ket{10,0}$ on $5\%$, giving the same fidelity as the phase-shifted source. The extra counts raise the expected number of flags in a full batch to about $293$, so nearly all runs stop at Stage~1. Every run that reaches Stage~2 passes because the full-space test accepts both the target and the Fock state above the cutoff. 

These examples show how the two tests address different errors at the same fidelity. Both final acceptance probabilities are below $\delta$, as required by Thm.~\ref{thm:two-stage}. Appx.~\ref{app:gaussian-error-acceptance} gives the calculations.

\section{Applications to Non-Gaussian States}\label{sec:non-gaussian-applications}

For pure Gaussian targets, local preprocessing in Sec.~\ref{sec:gaussian-applications} yields thermal marginals with geometric tails and a cutoff \emph{logarithmic} in $1/\varepsilon$. Non-Gaussian targets with identical first and second moments can have different photon tails. A known finite mean photon number gives a sufficient cutoff of order at most $1/\varepsilon$ via Eq.~\eqref{eq:target-markov}. We choose the Stage~1 cutoff using matched counts for photon subtraction and addition, photon statistics of coherent superpositions, finite NOON support, or moment and envelope bounds. In Stage~2, projected Schmidt decompositions give one-way tests for the squeezed states and GKP Bell pairs. We use logical qubit tests for coherent superpositions and NOON states, local projectors for single-mode targets, and a DV verification protocol for the hypergraph state.

\subsection{Symmetric photon subtraction from a TMSV}\label{sec:photon-subtraction}

\paragraph{Target state.}
Let $\lambda=\tanh^2r\in(0,1)$ and let $\ell\geq1$ be an integer.
For $j=A,B$, let $a_j$ denote the annihilation operator of mode $j$. In its local Fock basis, $a_j\ket{0}=0$ and $a_j\ket{n}=\sqrt{n}\ket{n-1}$ for $n\geq1$.
Applying $a_A^\ell a_B^\ell$ to the TMSV in Eq.~\eqref{eq:tmsv} and normalizing gives the symmetrically photon-subtracted family~\cite{NavarreteBenlloch2012Enhancing}. Writing $\binom{n+\ell}{\ell}$ for the binomial coefficient and $Z_\ell(\lambda)$ for the normalization factor, we have
\begin{align}
\ket{\psi_{\rm ps}^{\left(\ell\right)}}=
\frac1{\sqrt{Z_\ell\left(\lambda\right)}}\sum_{n=0}^{\infty}
\binom{n+\ell}{\ell}\lambda^{n/2}\ket{n,n},
\label{eq:ps-family}
\end{align}
where the normalization has a closed form with a polynomial numerator,
\begin{align}
Z_\ell\left(\lambda\right)
=\sum_{n=0}^{\infty}\binom{n+\ell}{\ell}^{\!2}\lambda^n
=\frac{\sum_{j=0}^{\ell}\binom{\ell}{j}^{\!2}\lambda^j}
{\left(1-\lambda\right)^{2\ell+1}}.
\label{eq:photon-normalization}
\end{align}
The case $\ell=1$, one photon removed from each mode, is
\begin{align}
\ket{\psi_{\rm ps}}=
\frac{\left(1-\lambda\right)^{3/2}}{\sqrt{1+\lambda}}
\sum_{n=0}^{\infty}\left(n+1\right)\lambda^{n/2}\ket{n,n}.
\label{eq:ps}
\end{align}
These targets are entangled and non-Gaussian for $\ell\geq1$ and $\lambda>0$. 
Their total mean photon number satisfies
\begin{align}
  E_{\rm ps}^{\left(\ell\right)}=2\lambda\partial_\lambda\ln Z_\ell\left(\lambda\right)<\infty.
\end{align}

Photon subtraction supplies a heralded non-Gaussian operation for CV entanglement distillation and can improve teleportation with a squeezed entangled resource~\cite{Opatrny2000Teleportation,Wang2015PhotonTeleportation}. Such resource engineering complements Gaussian optical processing, whose extension to universal quantum computing requires non-Gaussian ingredients~\cite{Walschaers2021NonGaussian}.
Appx.~\ref{app:photon-subtraction} illustrates the preparation of this family with a heralded optical circuit.

\paragraph{Two-stage verification.}
For Stage~1, the matched photon counts of the target give the exact joint tail
\begin{align}
q_K\left(\psi_{\rm ps}^{\left(\ell\right)}\right)=\frac1{Z_\ell\left(\lambda\right)}
\sum_{n=K+1}^{\infty}\binom{n+\ell}{\ell}^{\!2}\lambda^n.
\label{eq:ps-tail}
\end{align}
For arbitrary $\ell$, any $\lambda<z<1$ gives the computable bound
\begin{align}
q_K\left(\psi_{\rm ps}^{\left(\ell\right)}\right)\leq
\left(\frac\lambda z\right)^{K+1}
\frac{Z_\ell\left(z\right)}{Z_\ell\left(\lambda\right)}.
\label{eq:ps-exponential-tail}
\end{align}
For example, set $z=(1+\lambda)/2$ and choose
\begin{align}
K=\max\left\{1,\left\lceil
\frac{\ln\left(32/\varepsilon\right)+\ln\left[Z_\ell\left(z\right)/Z_\ell\left(\lambda\right)\right]}
{\ln\left(z/\lambda\right)}\right\rceil-1\right\}.
\label{eq:ps-cutoff}
\end{align}
This guarantees $q_K(\psi_{\rm ps}^{(\ell)})\leq\varepsilon/32$ using the closed form for $Z_\ell$ in Eq.~\eqref{eq:photon-normalization}. An exact tail calculation may give a smaller positive cutoff. For fixed $\ell$, the sufficient cutoff is $\cO((1-\lambda)^{-1}\ln(1/\varepsilon))$ as squeezing increases.

For Stage~2, the normalized projected target has Schmidt probabilities
\begin{align}
p_n^{\left(\ell,K\right)}=
\frac{\binom{n+\ell}{\ell}^{\!2}\lambda^n}
{\sum_{j=0}^K\binom{j+\ell}{\ell}^{\!2}\lambda^j},
\quad 0\leq n\leq K.
\label{eq:ps-schmidt}
\end{align}
Put $p_{\max}^{(\ell,K)}=\max_np_n^{(\ell,K)}$. Use the one-way bipartite verification protocol with the pass effect in Eq.~\eqref{eq:tmsv-finite-verifier}, replacing its Schmidt probabilities by Eq.~\eqref{eq:ps-schmidt} and its mixing weight by $w_K=p_{\max}^{(\ell,K)}/(1+p_{\max}^{(\ell,K)})$~\cite{Yu2019Optimal,Li2019Efficient}. It has
\begin{align}
\nu_K\geq\frac1{1+p_{\max}^{\left(\ell,K\right)}}\geq\frac12.
\label{eq:ps-gap}
\end{align}
The first inequality is an equality for a nontrivial Schmidt support; a one-dimensional sector instead has gap one. This gives an explicit LOCC verification protocol in the abstract measurement model, including when the largest probability occurs away from $n=0$.

With the cutoff chosen from the subtraction tail, the largest projected Schmidt probability controls the Stage~2 bound. Cor.~\ref{cor:photon-subtraction} gives the resulting sample complexity. 
Appx.~\ref{app:photon-subtraction} 
proves the corollary and gives a closed form for the tail when one photon is subtracted from each mode.

\begin{corollary}
\label{cor:photon-subtraction}
For the two-mode squeezed target with $\ell$ photons subtracted from each mode, Prot.~\ref{prot:two-stage} accepts the target with probability at least $1-\delta$ and any $\varepsilon$-bad source with probability at most $\delta$. A sufficient number of source copies is
\begin{align}
N_{\rm ps}^{\left(\ell\right)}=\left\lceil
\frac{98+2p_{\max}^{\left(\ell,K\right)}}{\varepsilon}\ln\left(1/\delta\right)
\right\rceil+1.
\label{eq:ps-samples}
\end{align}
\end{corollary}

\subsection{Symmetric photon addition to a TMSV}\label{sec:photon-addition}

\paragraph{Target state.}
For $j=A,B$, the creation operator $a_j^\dagger$ is the adjoint of $a_j$ and satisfies $a_j^\dagger\ket{n}=\sqrt{n+1}\ket{n+1}$ for $n\geq0$.
Applying $a_A^{\dagger\ell}a_B^{\dagger\ell}$ to the same TMSV and normalizing gives the 
symmetrically photon-added family~\cite{NavarreteBenlloch2012Enhancing}
\begin{align}
\ket{\psi_{\rm pa}^{\left(\ell\right)}}=
\frac1{\sqrt{Z_\ell\left(\lambda\right)}}\sum_{n=0}^{\infty}
\binom{n+\ell}{\ell}\lambda^{n/2}\ket{n+\ell,n+\ell},
\label{eq:pa-family}
\end{align}
where $\ell\geq1$, $0<\lambda<1$, and $Z_\ell$ is given by Eq.~\eqref{eq:photon-normalization}. For $\ell=1$, Eq.~\eqref{eq:pa-family} has the amplitudes of Eq.~\eqref{eq:ps} on $\ket{n+1,n+1}$. These states are non-Gaussian and entangled, with finite total mean photon number
\begin{align}
E_{\rm pa}^{\left(\ell\right)}=E_{\rm ps}^{\left(\ell\right)}+2\ell.
\label{eq:pa-energy}
\end{align}
At $\lambda=0$ their well-defined boundary state is the separable Fock product $\ket{\ell,\ell}$.

Photon addition supplies a complementary heralded route to non-Gaussian optical resource engineering and entanglement concentration~\cite{NavarreteBenlloch2012Enhancing}. Its relation to subtraction is especially transparent here: 
the two symmetric families have identical Schmidt probabilities at fixed $(\ell,\lambda)$, so the pure states have the same entanglement. Their quadrature correlations and teleportation performance can differ~\cite{Wang2015PhotonTeleportation}. Thus they provide distinct non-Gaussian targets relevant to CV quantum information despite this shared entanglement spectrum.
Appx.~\ref{app:photon-addition} illustrates the preparation of this family with a heralded optical circuit.

\paragraph{Two-stage verification.}
For Stage~1, the shift in photon number gives the exact tail relation
\begin{align}
q_K(\psi_{\rm pa}^{(\ell)})=
\begin{cases}
1,&K<\ell,\\
q_{K-\ell}(\psi_{\rm ps}^{(\ell)}),&K\geq\ell.
\end{cases}
\label{eq:pa-tail}
\end{align}
Choose a nonnegative auxiliary cutoff $J$ for the subtraction tail using the ceiling term minus one in Eq.~\eqref{eq:ps-cutoff} or an exact tail calculation. The physical cutoff $K=J+\ell\geq1$ then satisfies $q_K(\psi_{\rm pa}^{(\ell)})\leq\varepsilon/32$.

For Stage~2, let $p_n^{(\ell,J)}$ be the subtraction Schmidt probabilities in Eq.~\eqref{eq:ps-schmidt} evaluated at cutoff $J$, and let $p_{\max}^{(\ell,J)}$ be their maximum. The projected target is
\begin{align}
\ket{\psi_{{\rm pa},K}^{\left(\ell\right)}}=
\sum_{n=0}^{J}\sqrt{p_n^{\left(\ell,J\right)}}\ket{n+\ell,n+\ell}.
\label{eq:pa-truncated}
\end{align}
Apply the full-space rule of Eq.~\eqref{eq:dv-round-outcome}, using the coherence-preserving cutoff measurement in Appx.~\ref{app:dv-implementation}. An above-cutoff outcome gives a pass. Within the cutoff, test membership in the local support $\operatorname{span}\{\ket\ell,\ldots,\ket K\}$ coherently. The finite-sector test rejects outside those supports and uses the same one-way verification protocol in the shifted Schmidt bases inside them. This gives
\begin{align}
\nu_K\geq\frac1{1+p_{\max}^{\left(\ell,J\right)}}\geq\frac12.
\label{eq:pa-gap}
\end{align}

The cutoff shift accounts for the added photons and lets the corresponding subtraction Schmidt probabilities determine the Stage~2 bound. Cor.~\ref{cor:photon-addition} gives the resulting sample complexity.
Appx.~\ref{app:photon-addition} proves the corollary and gives a closed form for the tail when one photon is added to each mode.

\begin{corollary}
\label{cor:photon-addition}
For the two-mode squeezed target with $\ell$ photons added to each mode, Prot.~\ref{prot:two-stage} accepts the target with probability at least $1-\delta$ and any $\varepsilon$-bad source with probability at most $\delta$. A sufficient number of source copies is
\begin{align}
N_{\rm pa}^{\left(\ell\right)}=\left\lceil
\frac{98+2p_{\max}^{\left(\ell,J\right)}}{\varepsilon}\ln\left(1/\delta\right)
\right\rceil+1.
\label{eq:pa-samples}
\end{align}
\end{corollary}

\subsection{Multimode entangled coherent states}\label{sec:entangled-coherent-states}

\paragraph{Target state.}
Write $\mathbb C$ and $\mathbb R$ for the sets of complex and real numbers, respectively.
For $z\in\mathbb C$, write $\operatorname{Re}(z)$ for its real part.
For $\boldsymbol z=(z_1,\ldots,z_m)\in\mathbb C^m$, its Euclidean norm is $\lVert\boldsymbol z\rVert=\sqrt{\sum_{j=1}^m\vert z_j\vert^2}$.
Let $\boldsymbol\alpha,\boldsymbol\beta\in\mathbb C^m$ be vectors of known coherent amplitudes, and write $\ket{\boldsymbol\alpha}=\bigotimes_{j=1}^m\ket{\alpha_j}$ for the product of local coherent states, and similarly for $\ket{\boldsymbol\beta}$. With known relative phase $\varphi\in\mathbb R$, the balanced two-branch family of entangled coherent states (ECSs) is~\cite{Sanders2012ECS}
\begin{subequations}\label{eq:ecs-family}
\begin{align}
\ket{\Psi_{\rm ECS}}&=
\frac{\ket{\boldsymbol\alpha}+e^{i\varphi}\ket{\boldsymbol\beta}}{\sqrt{\mathcal Z}},\\
\mathcal Z&=2+2\operatorname{Re}(e^{i\varphi}\braket{\boldsymbol\alpha}{\boldsymbol\beta})>0,\\
\braket{\boldsymbol\alpha}{\boldsymbol\beta}
&=\exp\!\left[-\frac{\left\lVert\boldsymbol\alpha\right\rVert^2+\left\lVert\boldsymbol\beta\right\rVert^2}{2}
+\sum_{j=1}^m\alpha_j^*\beta_j\right].
\end{align}
\end{subequations}
If $m\geq2$ and $\alpha_j\neq\beta_j$ for every mode, the two local coherent states are linearly independent at every party, and the state is entangled across every bipartition. The two-mode target $\ket{\alpha,0}+\ket{0,\beta}$ and the GHZ-like target $\bigotimes_j\ket{\alpha_j}+\ket{0}^{\otimes m}$ studied in Ref.~\cite{Liu2021CVQSV} are members of Eq.~\eqref{eq:ecs-family}, with normalization understood. A local displacement on the second mode maps the former to $\ket{\alpha,-\beta}+\ket{0,0}$, so the two-mode example is locally equivalent to a member of the latter family.

Choosing $\boldsymbol\beta=-\boldsymbol\alpha$ gives an entangled state in the tensor product of local cat-qubit spaces $\operatorname{span}\{\ket{\alpha_j},\ket{-\alpha_j}\}$. These local spaces are used for bosonic encodings in propagating light~\cite{konno2024logical}.

\paragraph{Two-stage verification.}
For Stage~1, let $\gamma,\zeta\in\mathbb C$ denote single-mode coherent amplitudes. The local cutoff probability and overlap are
\begin{subequations}\label{eq:ecs-cutoff-functions}
\begin{align}
f_K\left(\gamma\right)&=e^{-\left\lvert\gamma\right\rvert^2}\sum_{n=0}^K\frac{\left\lvert\gamma\right\rvert^{2n}}{n!},\\
h_K\left(\gamma,\zeta\right)&=
e^{-\left(\left\lvert\gamma\right\rvert^2+\left\lvert\zeta\right\rvert^2\right)/2}
\sum_{n=0}^K\frac{\left(\gamma^*\zeta\right)^n}{n!}.
\end{align}
\end{subequations}
Factorization of $P_K$ gives the interference term $\cI_K$ between the two branches and the exact target tail
\begin{subequations}
\begin{align}
\cI_K&=2\operatorname{Re}\!\left(e^{i\varphi}
\prod_{j=1}^m h_K\left(\alpha_j,\beta_j\right)\right),\\
q_K\left(\Psi_{\rm ECS}\right)&=
1-\frac{\prod_j f_K\left(\alpha_j\right)+\prod_j f_K\left(\beta_j\right)+\cI_K}{\mathcal Z}.
\label{eq:ecs-tail}
\end{align}
\end{subequations}
Choose the least positive $K$ satisfying $q_K(\Psi_{\rm ECS})\leq\varepsilon/32$ from Eq.~\eqref{eq:ecs-tail}.

For Stage~2, let $\ket{\gamma;K}=e^{-\lvert\gamma\rvert^2/2}\sum_{n=0}^K\gamma^n\ket n/\sqrt{n!}$ be an unnormalized truncated coherent state. The projected target is
\begin{align}
\ket{\Psi_{{\rm ECS},K}}=
\frac{\bigotimes_j\ket{\alpha_j;K}
+e^{i\varphi}\bigotimes_j\ket{\beta_j;K}}
{\sqrt{\mathcal Z\left(1-q_K\left(\Psi_{\rm ECS}\right)\right)}}.
\label{eq:ecs-truncated}
\end{align}
For $K\geq1$, the local support spanned by $\ket{\alpha_j;K}$ and $\ket{\beta_j;K}$ has dimension one when $\alpha_j=\beta_j$ and two otherwise. The ratio of the one-photon and vacuum amplitudes distinguishes the two truncated states for unequal amplitudes. Use the full-space rule of Eq.~\eqref{eq:dv-round-outcome}. After a coherent cutoff check in Appx.~\ref{app:dv-implementation}, any above-cutoff outcome gives a pass. For copies within the cutoff, test membership in each support coherently and reject a failed support check. A one-dimensional factor is checked by its local projector. Apply the adaptive strategy of Li and Zhu~\cite{Li2026Universal} to the remaining qubits when there are at least two. For a product target, local target projectors suffice. There are at most $m$ qubit factors, so Eq.~\eqref{eq:graph-general-gap} gives the certified bound $\nu_K\geq2^{1-m}$.

With the cutoff fixed by the exact tail, Cor.~\ref{cor:ecs-verification} expresses the sample complexity in terms of the in-sector gap $\nu_K$. Appx.~\ref{app:ecs-proof} proves the bound.

\begin{corollary}
\label{cor:ecs-verification}
For the multimode entangled coherent target $\ket{\Psi_{\rm ECS}}$ in Eq.~\eqref{eq:ecs-family}, Prot.~\ref{prot:two-stage} accepts the target with probability at least $1-\delta$ and any $\varepsilon$-bad source with probability at most $\delta$. A sufficient number of source copies is
\begin{align}
N_{\rm ECS}=\left\lceil\frac{96+2/\nu_K}{\varepsilon}\ln\left(1/\delta\right)\right\rceil+1.
\label{eq:ecs-samples}
\end{align}
\end{corollary}

\begin{figure}[t]
\centering
\includegraphics[width=\columnwidth]{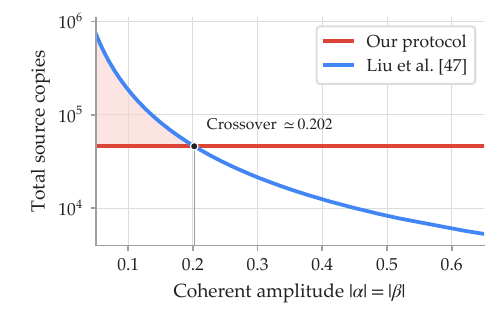}
\caption{\raggedright\textbf{Total sample complexity for two-mode entangled coherent states.}
The target is proportional to $\ket{\alpha,0}+\ket{0,\beta}$ with $\lvert\alpha\rvert=\lvert\beta\rvert$. Both curves use $\varepsilon=\delta=0.01$ and the specified ideal local measurements. Red shows a sufficient total for Prot.~\ref{prot:two-stage} from the certified gap bound. Blue shows the exact worst-case copy count for the settings of Ref.~\cite{Liu2021CVQSV}, with optimized setting probabilities, independent draws, and an all-pass decision.}
\label{fig:ecs-copy-comparison}
\end{figure}

\Needspace{9\baselineskip}
\paragraph{Comparison with~\cite{Liu2021CVQSV}.}
Both protocols apply to the two-mode targets proportional to $\ket{\alpha,0}+\ket{0,\beta}$ and the multimode GHZ-like targets proportional to $\bigotimes_{j=1}^m\ket{\alpha_j}+\ket{0}^{\otimes m}$.
Liu \emph{et al.} achieve perfect completeness for these shared targets using only local displacements and photon counting.
At common amplitude $0.1$, our sufficient copy budget is about one quarter of the direct test's optimized worst-case requirement. The curves in Fig.~\ref{fig:ecs-copy-comparison} cross near amplitude $0.202$. Above the crossing, the direct test's requirement is below our sufficient two-stage budget, and at amplitude $0.515$ it is below our Stage~1 allocation. At sufficiently small common amplitudes, the sufficient two-stage budget also falls below the direct test's worst-case requirement for GHZ-like targets.
Our DV test incorporates any known relative phase in Eq.~\eqref{eq:ecs-family}, so it also covers targets that yield both displaced parity outcomes.
With trusted cutoff-preserving basis rotations, the detectors need to resolve counts through $K$ and flag larger counts. The Fock tail test uses those flags, and our copy guarantee includes the flagged rounds. Liu \emph{et al.} discard saturated records in their finite-resolution parity measurements.
Appx.~\ref{app:ecs-liu-comparison} gives the analysis.

\subsection{Balanced multimode NOON states}\label{sec:noon}

\paragraph{Target state.}
Let $\ket{s_j}$ denote the Fock state with all $s\geq1$ photons in mode $j$ and vacuum in the other $m-1$ modes. The balanced $m$-mode NOON state is
\begin{align}
\ket{\Psi_{m,s}}=\frac{1}{\sqrt m}\sum_{j=1}^{m}\ket{s_j},
\qquad m\geq2.
\label{eq:noon-state}
\end{align}
A phase shift $\theta_j$ in path $j$ gives that branch the phase $s\theta_j$, making these superpositions useful probes of relative phases in multiport interferometers~\cite{Humphreys2013Phases}. To share entanglement between two chosen paths, measure the other paths in the number basis. With probability $2/m$ they all give vacuum, leaving an encoded Bell pair for teleportation or a remote controlled-NOT gate~\cite{Eisert2000Nonlocal}. A heralded optical experiment has prepared the $m=3$, $s=2$ state~\cite{Singh2026NOON}.

\paragraph{Two-stage verification.}
For Stage~1, the least admissible cutoff is exactly $K=s$. Every target copy therefore passes with certainty.

For Stage~2, identify $\ket0$ and $\ket s$ in each mode with the basis states of a logical qubit. The target in Eq.~\eqref{eq:noon-state} is the $m$-qubit $W$ state with one logical excitation. Apply the full-space rule of Eq.~\eqref{eq:dv-round-outcome} with the coherent cutoff check in Appx.~\ref{app:dv-implementation}. An above-cutoff outcome gives a pass. Within the cutoff, each party tests membership in the span of $\ket0$ and $\ket s$ while preserving coherence. A failed logical support check rejects the copy. Otherwise the parties apply the adaptive $W$ test for $m\geq3$ or the three Pauli parity tests for a Bell pair when $m=2$~\cite{Liu2019Dicke}. The resulting gaps are
\begin{align}
\nu_K=
\begin{cases}
2/3, & m=2,\\
1/3, & m=3,\\
1/\left(m-1\right), & m\geq4.
\end{cases}
\label{eq:noon-gap}
\end{align}

At $K=s$, the exact target passes both stages with certainty. Cor.~\ref{cor:noon} gives the sample complexity using the gap of the logical test. Appx.~\ref{app:noon} specifies the local measurement settings and proves the corollary.

\begin{corollary}\label{cor:noon}
Prot.~\ref{prot:two-stage} accepts $\ket{\Psi_{m,s}}$ with certainty and any $\varepsilon$-bad source with probability at most $\delta$. With $\nu_K$ from Eq.~\eqref{eq:noon-gap}, a sufficient number of source copies is
\begin{align}
N=\left\lceil\frac{96+2/\nu_K}{\varepsilon}\ln\left(1/\delta\right)\right\rceil+1.
\label{eq:noon-samples}
\end{align}
\end{corollary}

\subsection{Finite-energy GKP Bell pairs}\label{sec:gkp-bell}

\paragraph{Target state.}
Use dimensionless position and momentum quadratures $\hat q$ and $\hat p$ with $[\hat q,\hat p]=i$ and vacuum position variance $1/2$. Let $\ket{\mu}_{\rm id}$ be the ideal square-lattice comb for logical label $\mu\in\{0,1\}$. With $\ket q$ the position eigenket, $\delta_{\rm D}$ the Dirac delta, and $\mathbb Z$ the set of integers, its wave function is $\braket{q}{\mu}_{\rm id}=\sum_{s\in\mathbb Z}\delta_{\rm D}(q-(2s+\mu)\sqrt\pi)$.
Apply the energy envelope $D_t=e^{-t(\hat n_A+\hat n_B)}$, $t>0$, to the sum of the two product combs. For $\mu,\nu\in\{0,1\}$, let $G_{\mu\nu}(t)$ be the overlap of the single-mode filtered combs $e^{-t\hat n}\ket{\mu}_{\rm id}$ and $e^{-t\hat n}\ket{\nu}_{\rm id}$. Set $C_t=[\pi(1-e^{-4t})]^{-1/2}$ and, for real $x,y$, $H_t(x,y)=\tanh t\,(x+y)^2+\coth t\,(x-y)^2$. The real overlaps and the squared norm of the filtered sum are
\begin{align}
G_{\mu\nu}\left(t\right)
&=C_t\sum_{s,u\in\mathbb Z}e^{-\pi H_t\left(s+\mu/2,u+\nu/2\right)},
\label{eq:gkp-bell-overlaps}\\
Z_{\rm B}\left(t\right)
&=G_{00}^2\left(t\right)+G_{11}^2\left(t\right)+2G_{01}^2\left(t\right).
\label{eq:gkp-bell-normalization}
\end{align}
The normalized two-oscillator target is
\begin{align}
\ket{\Phi_t}
=\frac{D_t}{\sqrt{Z_{\rm B}\left(t\right)}}
\sum_{\mu=0}^{1}\ket{\mu}_{\rm id}\otimes\ket{\mu}_{\rm id}.
\label{eq:gkp-bell-state}
\end{align}

GKP Bell pairs supply logical entanglement for optical quantum repeaters~\cite{Fukui2021Repeater}. A trapped-ion experiment has both implemented an entangling gate on two GKP qubits and directly prepared a logical Bell state in two motional modes~\cite{Matsos2025Gates}. The grid correlations also give a displacement signal. Let $q_A$ and $q_B$ be the position outcomes in the two modes. In the ideal Bell state, $q_A-q_B$ is an integer multiple of $2\sqrt\pi$, so a relative position shift displaces the peaks of the difference distribution. The finite-energy envelope sets their widths.

\paragraph{Two-stage verification.}
For Stage~1, use the energy envelope to choose the cutoff. If either count exceeds $K$, their sum is at least $K+1$, so
\begin{align}
q_K\left(\Phi_t\right)
\leq e^{-t\left(K+1\right)}
\frac{Z_{\rm B}\left(t/2\right)}{Z_{\rm B}\left(t\right)}.
\label{eq:gkp-bell-tail}
\end{align}
At fixed $t$, a cutoff logarithmic in $1/\varepsilon$ makes this bound at most $\varepsilon/32$.

For Stage~2, project both filtered combs in each mode onto levels $0,\ldots,K$. Their Bell superposition has Schmidt rank at most two. The filtered combs can overlap, so compute the Schmidt bases and probabilities of the normalized projected target. With $p_{\max}$ its largest Schmidt probability, the one-way bipartite test in these bases gives $\nu_K\geq1/(1+p_{\max})\geq1/2$~\cite{Yu2019Optimal,Li2019Efficient}.

The energy envelope determines a sufficient cutoff, and the one-way test has a gap bound independent of $K$. Cor.~\ref{cor:gkp-bell} gives the resulting sample complexity. Appx.~\ref{app:gkp-bell} supplies the Fock coefficients needed for both stages and proves the corollary.

\begin{corollary}\label{cor:gkp-bell}
For every $t>0$, Prot.~\ref{prot:two-stage} accepts $\ket{\Phi_t}$ with probability at least $1-\delta$ and any $\varepsilon$-bad source with probability at most $\delta$. A sufficient number of source copies is
\begin{align}
N=\left\lceil\frac{100}{\varepsilon}\ln\left(1/\delta\right)\right\rceil+1.
\label{eq:gkp-bell-samples}
\end{align}
\end{corollary}

\subsection{Finitely squeezed three-mode CV hypergraph states}\label{sec:cv-hypergraph}

\paragraph{Target state.}
Let $\hat q_j$ and $\hat p_j$ be the position and momentum quadrature operators of mode $j$, with $[\hat q_j,\hat p_k]=i\delta_{jk}$, where $\delta_{jk}$ is the Kronecker delta. For finite input squeezing $r\geq0$, write $\ket{0;r}_p$ for the momentum-squeezed vacuum with position variance $e^{2r}/2$ and momentum variance $e^{-2r}/2$.
A single hyperedge of real strength $\gamma\ne0$ couples three such vacua through a cubic phase~\cite{Moore2019Hypergraph},
\begin{align}
\ket{H_{\gamma,r}}
=e^{i\gamma\hat q_1\hat q_2\hat q_3}\ket{0;r}_p^{\otimes3}.
\label{eq:cv-hypergraph-state}
\end{align}

This interaction makes the state non-Gaussian. Three-mode hyperedges enable non-Gaussian gates in measurement-based computation~\cite{Moore2019Hypergraph}. Measuring $q_3=x$ leaves modes 1 and 2 coupled by $e^{i\gamma x\hat q_1\hat q_2}$, so any nonzero outcome localizes entanglement between the two remaining parties. The coupling itself can be estimated by measuring $q_1,q_2,p_3$ locally, since their mean product is $\gamma e^{4r}/4$. A related superconducting-cavity experiment used three-photon down-conversion to produce three microwave modes and detected their genuine non-Gaussian entanglement through three-mode correlations~\cite{JarvisFrain2026Tripartite}.

\paragraph{Two-stage verification.}
For Stage~1, the target's total mean photon number $E_{\gamma,r}$ bounds the tail,
\begin{align}
E_{\gamma,r}&=3\sinh^2r+\frac{3\gamma^2 e^{4r}}{8},
\label{eq:cv-hypergraph-energy}\\
q_K\left(H_{\gamma,r}\right)&\leq\frac{E_{\gamma,r}}{K+1}.
\label{eq:cv-hypergraph-tail}
\end{align}
Taking $K+1\geq32E_{\gamma,r}/\varepsilon$ suffices.

For Stage~2, expand the known target in Eq.~\eqref{eq:cv-hypergraph-state} in the Fock basis through $K$ in each mode and normalize. Apply the adaptive LOCC strategy of Sec.~\ref{sec:multimode-Gaussian-states} to this projected target. Its four equally likely settings give $\nu_K\geq1/4$ for all $K$~\cite{Li2026Universal}.

The target's mean photon number sets a sufficient cutoff. The gap bound for the tripartite test then gives the sample complexity in Cor.~\ref{cor:cv-hypergraph}. Appx.~\ref{app:cv-hypergraph} derives the mean photon number and proves the corollary.

\begin{corollary}\label{cor:cv-hypergraph}
Prot.~\ref{prot:two-stage} accepts $\ket{H_{\gamma,r}}$ with probability at least $1-\delta$ and any $\varepsilon$-bad source with probability at most $\delta$. A sufficient number of source copies is
\begin{align}
N=\left\lceil\frac{104}{\varepsilon}\ln\left(1/\delta\right)\right\rceil+1.
\label{eq:cv-hypergraph-samples}
\end{align}
\end{corollary}

\subsection{Single-mode non-Gaussian resources}\label{sec:single-mode-pure-states}

\paragraph{Target state.}
We apply Prot.~\ref{prot:two-stage} to three representative single-mode non-Gaussian resources.

\emph{Cat states.} For $\alpha\ne0$ and $\mu=\lvert\alpha\rvert^2$, the even and odd cat states are
\begin{align}
\ket{C_\pm(\alpha)}=
\frac{\ket\alpha\pm\ket{-\alpha}}{\sqrt{2\left(1\pm e^{-2\mu}\right)}}.
\label{eq:single-cat}
\end{align}
Their span encodes a bosonic qubit. Squeezed cat states provide a route to preparing logical states in propagating light~\cite{konno2024logical}.

\emph{Cubic phase states.} Use single-mode position and momentum operators $\hat q$ and $\hat p$ in the convention $[\hat q,\hat p]=i$. Applying a cubic phase of real strength $\gamma\ne0$ to a centered squeezed vacuum with position standard deviation $\sigma>0$ yields $\ket{\psi_{\gamma,\sigma}}$. Its normalized wave function in the position coordinate $q$ is~\cite{Brunelli2019Reservoir}
\begin{align}
\psi_{\gamma,\sigma}\left(q\right)=
\frac{\exp\left[-q^2/\left(4\sigma^2\right)+i\gamma q^3\right]}{\left(2\pi\sigma^2\right)^{1/4}}.
\label{eq:single-cubic}
\end{align}
This state supplies the non-Gaussian resource for implementing a cubic phase gate in CV quantum computation~\cite{Walschaers2021NonGaussian}.

\emph{GKP states.} For logical label $\mu_L\in\{0,1\}$, use the ideal square-lattice comb $\ket{\mu_L}_{\rm id}$ defined in Sec.~\ref{sec:gkp-bell}.
For $t>0$, let $Z_{\mu_L}(t)$ be the squared norm of the comb after applying the energy envelope $e^{-t\hat n}$. The normalized finite-energy state is~\cite{Sivak2022Control}
\begin{align}
\ket{G_{\mu_L,t}}=Z_{\mu_L}\left(t\right)^{-1/2}
e^{-t\hat n}\ket{\mu_L}_{\rm id}.
\label{eq:single-gkp}
\end{align}
The grid encodes qubits that can be protected against displacement errors, and integrated photonic sources have generated optical GKP qubits~\cite{larsen2025gkp}.

\paragraph{Two-stage verification.}
For a single-mode known target $\ket{\psi}$ as above, choose a positive Stage~1 cutoff $K$ so that a certified tail bound $\tau_K\geq q_K(\psi)$ is at most $\varepsilon/32$. When the mean photon number $E_\psi=\bra{\psi}\hat n\ket{\psi}$ is finite, a general choice is $\tau_K=E_\psi/(K+1)$. For a squeezed cat, apply the inverse squeezing to every copy before both stages and use the cat photon statistics in that frame. Cat states admit an exact tail from the Poisson weights of the corresponding parity,
\begin{align}
q_K(C_\pm(\alpha))
=\frac{e^{-\mu}}{1\pm e^{-2\mu}}
\sum_{n>K}\frac{\left[1\pm\left(-1\right)^n\right]\mu^n}{n!}.
\label{eq:single-cat-tail}
\end{align}
For a normalized state $\ket\phi$, write $\langle A\rangle_\phi=\bra\phi A\ket\phi$ for the expectation of an operator $A$. For the cubic phase state, use a nonempty finite set $\mathcal J$ of positive integer moment orders to bound the tail,
\begin{align}
q_K(\psi_{\gamma,\sigma})\leq\min_{j\in\mathcal J}\frac{M_j}{(K+1)^j},\qquad 
M_j=\langle\hat n^j\rangle_{\psi_{\gamma,\sigma}}.
\label{eq:single-moment-cutoff}
\end{align}
For the GKP state, the envelope gives, for $0<h<t$,
\begin{align}
q_K\left(G_{\mu_L,t}\right)\leq e^{-2h\left(K+1\right)}\frac{Z_{\mu_L}\left(t-h\right)}{Z_{\mu_L}\left(t\right)}.
\label{eq:single-gkp-tail}
\end{align}
For example, $h=t/2$ gives a cutoff logarithmic in $1/\varepsilon$ at fixed $t$. Appx.~\ref{app:single-mode-practical} gives the Fock coefficients and the evaluation of the moments and normalization.

For Stage~2, normalize $P_K\ket\psi$ to obtain $\ket{\psi_K}$. A rank-one projector is a local test for a single mode, so the test has a unit gap $\nu_K=1$, and
\begin{align}
\Omega_K=\proj{\psi_K},\quad
\Omega_K^+=\proj{\psi_K}+P_K^{\perp}.
\label{eq:single-mode-projector}
\end{align}
One realization of this unit-gap test assumes access to a trusted local basis rotation followed by PNRD.

Appx.~\ref{app:dv-implementation} also gives a single-mode product projector strategy using coherent references and photon counting.
Each round samples one vector uniformly from an orthonormal basis of the $K$-dimensional subspace orthogonal to $\ket{\psi_K}$ within $P_K\cH$ and rejects only on that vector's prescribed photon count record.
The smallest probability of obtaining a prescribed record when the input is its corresponding rejection vector, divided by $K$, gives the optical gap $\nu_K^{\rm opt}$ used to set the Stage~2 sample complexity in Prot.~\ref{prot:two-stage}.

The three resources illustrate how to choose the Stage~1 cutoff from the target's photon statistics.
The unit-gap Stage~2 measurement gives the common sample complexity bound in Cor.~\ref{cor:single-mode}, proved in Appx.~\ref{app:single-mode-proof}.

\begin{corollary}
\label{cor:single-mode}
Let $\ket{\psi}$ be a known pure single-mode non-Gaussian target with finite mean photon number, such as a cat, cubic phase, or finite-energy GKP state as described above.
With the Stage~2 projector in Eq.~\eqref{eq:single-mode-projector}, Prot.~\ref{prot:two-stage} accepts the exact target with probability at least $1-\delta$ and any $\varepsilon$-bad source with probability at most $\delta$. A sufficient number of source copies is
\begin{align}
N_{\rm single}=\left\lceil\frac{98}{\varepsilon}\ln\left(1/\delta\right)\right\rceil+1.
\label{eq:single-mode-samples}
\end{align}
\end{corollary}

\section{Conclusions}\label{sec:conclusions}

We have established a two-stage LOCC protocol for any known multimode pure CV target using independent, identically distributed source copies and trusted local measurements.
The protocol rests on the observation that we can choose a common cutoff from the target's photon statistics to bound all its counts with arbitrarily high probability.
A source at or below the fidelity threshold must then have either excessive weight above this cutoff or detectable weight orthogonal to the projected target in the subspace it defines.
The Fock tail test and DV state verification address these two possibilities, respectively, giving a full-state fidelity guarantee without an energy bound on the source.
At fixed confidence and mode number, the sample complexity scales inversely with the infidelity threshold if the ideal local DV gap stays bounded away from zero as the cutoff grows.
Applications to experimentally relevant Gaussian and non-Gaussian states demonstrate the protocol's versatility.

For multimode optical implementations, an open question is how the attainable verification gap depends on local phase references and the range of photon numbers the detectors can resolve.
Determining how these resources must scale with the cutoff 
would give quantitative criteria for preserving the inverse scaling.

\section*{Acknowledgments}
This work was supported by 
the National Natural Science Foundation of China (Grant No. 12504584),
the Quantum Science and Technology-National Science and Technology Major Project (Grant No. 2025ZD0300300), and 
the Innovation Research Foundation of NUDT (Grant No. 202501-YJRC-ZZ-004).

\makeatletter
\newcommand{\appendixtitle}[1]{\gdef\@title{#1}}
\makeatother

\makeatletter%
\newcommand{\appendixmaketitle}{%
\begin{center}%
\vspace{0.4in}%
{\large \@title \par}%
\end{center}%
\par%
}%
\makeatother%

\makeatletter
\newcommand{\appendixtableofcontents}{%
\begingroup
\linespread{0.95}\selectfont
\setcounter{tocdepth}{4}%
\@starttoc{atoc}%
\endgroup}
\makeatother

\newcommand{\appendixtocdivider}{%
\par\medskip
\noindent\hbox to \linewidth{%
\leaders\hbox{\rule[0.6ex]{1pt}{0.4pt}}\hfill
\hspace{0.8em}\textsc{Table of Contents}\hspace{0.8em}%
\leaders\hbox{\rule[0.6ex]{1pt}{0.4pt}}\hfill}%
\par}

\makeatletter
\newcounter{subsubsubsection}[subsubsection]
\renewcommand*{\p@subsubsubsection}{\p@subsubsection}
\renewcommand{\thesubsubsubsection}{\thesubsubsection.\arabic{subsubsubsection}}
\providecommand{\theHsubsubsubsection}{}
\renewcommand{\theHsubsubsubsection}{\theHsubsubsection.\arabic{subsubsubsection}}
\providecommand*{\toclevel@subsubsubsection}{4}
\newcommand{\subsubsubsection}[1]{%
\refstepcounter{subsubsubsection}%
\paragraph*{\thesubsubsubsection\space #1}}
\newcommand*\l@subsubsubsection[2]{\@dottedtocline{4}{4.4em}{3.0em}{#1}{#2}}
\makeatother

\newcommand{\appsection}[1]{%
\section{#1}%
\addcontentsline{atoc}{section}{\protect\numberline{\thesection}#1}}

\newcommand{\appsubsection}[1]{%
\subsection{#1}%
\addcontentsline{atoc}{subsection}{\protect\numberline{\thesubsection}#1}}

\newcommand{\appsubsubsection}[1]{%
\subsubsection{#1}%
\addcontentsline{atoc}{subsubsection}{\protect\numberline{\thesubsubsection}#1}}

\newcommand{\appproofheading}[1]{%
  \Needspace{6\baselineskip}%
  \appsubsubsection{Proof of #1}}

\newcommand{\appsubsubsubsection}[1]{%
\subsubsubsection{#1}}

\setcounter{secnumdepth}{4}

\appendix
\onecolumngrid
\newpage

\appendixtitle{\bf
Supplemental Material for\\``\thetitle''}
\appendixmaketitle
\vspace{0.1in}

\noindent
The appendices provide supporting details for the main text (MT) and are organized as follows.
Appx.~\ref{app:perfect-completeness} explains how perfect completeness restricts CV measurements.
Appx.~\ref{app:protocol} proves the guarantees of the two-stage protocol and describes the local optical measurements.
Appxs.~\ref{app:gaussian-applications} and~\ref{app:non-gaussian-applications} prove the corollaries for Gaussian and non-Gaussian targets, respectively, and give the calculations behind the numerical examples.

\appendixtocdivider
{%
\appendixtableofcontents%
}%

\appsection{Perfect Completeness and the CV Obstruction}\label{app:perfect-completeness}

Perfect completeness means that the exact target always passes. To detect an error, the verifier needs a set of outcomes with zero probability on the target and positive probability on the bad source. For a continuous readout, a single exact value has probability zero for every source with a density. The probability of the whole rejection set matters.
Concretely,
fix a setting $x$ with $p_x>0$ and let $z$ be the outcome record on one copy. Write $p_\psi(z)$ for the target density with respect to a reference measure $\mu$. For the continuous readouts below, $\mu$ is ordinary volume on the outcome space (Lebesgue measure). If the copy passes with probability $a(z)\in[0,1]$ upon observing $z$, the perfect completeness condition in Eq.~\eqref{eq:perfect-completeness} gives
\begin{align}
  \int\left[1-a\left(z\right)\right]p_\psi\left(z\right)\,d\mu\left(z\right)=0.
\end{align}
The integrand is nonnegative. If the target density is positive apart from a set of zero volume, the verifier must pass every outcome apart from a set of zero volume. Any source with a density on the same space passes with certainty. Relaxing perfect completeness allows the verifier to reject events that can occasionally occur on the target. The examples below show how perfect completeness constrains several common CV measurements.

\textbf{Homodyne measurement.}
Local homodyne detection measures one quadrature per mode. The joint outcome density is the corresponding marginal of the Wigner function~\cite{VogelRisken1989Quadrature}. For a finitely squeezed Gaussian target, this density is positive everywhere. Any rejection region with nonzero volume would sometimes reject the exact target. Perfect completeness therefore forces the test to pass every source. Other targets can give a useful rejection event in a particular setting. For example, if the target gives only nonnegative positions, the verifier can reject a negative position outcome and still have perfect completeness. A source with the same position probabilities, such as one with a different wavefunction phase, still passes that check.

\textbf{Heterodyne measurement.}
Local heterodyne detection gives one complex outcome per mode. Its outcome density is the Husimi function, which is real analytic for a pure target~\cite{ChabaudMehraban2022Holomorphic}. Because this density integrates to one, its zeros occupy zero volume~\cite{Mityagin2020ZeroSet}. Every source has a heterodyne outcome density, so that set of outcomes has probability zero for every source. A heterodyne test with perfect completeness thus passes every source.

\textbf{Photon counting measurement.}
Photon counting can reject some errors under perfect completeness. A two-mode squeezed vacuum always gives equal counts, so unequal counts rule it out. A source that randomly prepares $\ket{n,n}$ with the target's probabilities $(1-\lambda)\lambda^n$ for $n\geq0$ and $0<\lambda<1$ gives the same distribution of counts. The fidelity of this mixture with the target is $(1-\lambda)/(1+\lambda)$. Whenever the mixture qualifies as a bad source, a decision based on the counts accepts it as often as the target.

\textbf{Photon counting with passive optics.}
The same ambiguity persists with local passive interferometers and ancillas of definite photon number. Each interferometer conserves the total photon number of the signal and its ancillas. The ancillas and the photon-counting effects are invariant under a common phase shift. Hence the induced effect $B_j$ on signal mode $j$ obeys $e^{i\theta\hat n_j}B_je^{-i\theta\hat n_j}=B_j$ for every phase angle $\theta$ and is diagonal in the photon number basis. The invariance also holds when earlier counts guide later passive operations. Their outcomes therefore depend only on the joint photon number probabilities. A phase reference can make photon counting sensitive to coherence between different patterns.

\textbf{Perfect completeness and the verification gap.}
In infinite dimension, even a test that accepts only the target with certainty can have a zero verification gap.
For example, let $\{\ket{\phi_j}\}_{j\geq1}$ be an orthonormal basis of $\ket{\psi}^{\perp}$. The effect
\begin{align}
  \psi+\sum_{j\geq1}\frac{j}{j+1}\proj{\phi_j}
\end{align}
passes $\ket{\psi}$ with certainty and rejects $\ket{\phi_j}$ with probability $1/(j+1)$.
Because $\ket{\psi}^{\perp}$ is infinite dimensional, the denominator $j+1$ can grow without bound. The rejection probability therefore becomes arbitrarily small, and the gap in Eq.~\eqref{eq:general-gap} vanishes.
For any fixed number of copies, taking $j$ large enough makes the probability of passing all rounds arbitrarily close to one.

\appsection{Two-Stage Verification Protocol}\label{app:protocol}

Following Sec.~\ref{sec:protocol}, we give the supporting results for the Fock tail test and DV state verification, then derive bounds on the error probabilities and sample complexity of the combined protocol.

\appsubsection{Fock tail test}\label{app:cutoff-selection}\label{app:mean-energy}

Any strictly positive upper bound on the tail probability alone permits arbitrarily large total mean photon numbers. At the endpoint $q_K(\rho)=0$, the state is supported in $P_K\cH$ and its total mean photon number is at most $mK$.
To see the strictly positive case, take a target with finite total mean photon number $E_\psi$ and tail probability strictly below the desired bound. For $n>K$, let $\rho_{p,n}=(1-p)\psi+p\proj{n,0,\ldots,0}$ with $0<p<1$. The tail is $(1-p)q_K(\psi)+p$, so a sufficiently small fixed $p$ meets the bound for every $n$. The total mean photon number $(1-p)E_\psi+pn$ diverges as $n\to\infty$.
Any test on $N$ copies that accepts $\psi^{\otimes N}$ with probability at least $1-\delta_{\rm G}$ accepts $\rho_{p,n}^{\otimes N}$ with probability at least $(1-p)^N(1-\delta_{\rm G})$, because the all-target component has weight $(1-p)^N$. At fixed $N$, taking $p$ small makes this lower bound arbitrarily close to $1-\delta_{\rm G}$ for every $n$. The FTT estimates the expectation of the bounded effect $P_K^\perp$, which is the tail probability itself.

\appproofheading{Lem.~\ref{lem:ftt}}\label{app:binomial}

Let $q=q_K(\rho)$ and write $\Pr_p$ for probability under the binomial law $X\sim\mathrm{Bin}(N_1,p)$. For $q\leq\varepsilon/32$, stochastic monotonicity and the multiplicative Chernoff upper tail inequality give
\begin{align}
{\Pr}_q\left(X/N_1>\varepsilon/16\right)
\leq{\Pr}_{\varepsilon/32}\left(X\geq N_1\varepsilon/16\right)
\leq e^{-N_1\varepsilon/96}.
\label{eq:app-honest}
\end{align}
The final step uses $\Pr(X\geq2\mu)\leq e^{-\mu/3}$ with $\mu=N_1\varepsilon/32$. Event inclusion handles noninteger thresholds. For $q\geq\varepsilon/8$,
\begin{align}
{\Pr}_q\left(X/N_1\leq\varepsilon/16\right)
\leq{\Pr}_{\varepsilon/8}\left(X\leq N_1\varepsilon/16\right)
\leq e^{-N_1\varepsilon/64},
\label{eq:app-bad-tail}
\end{align}
where the final step uses $\Pr(X\leq\mu/2)\leq e^{-\mu/8}$ with $\mu=N_1\varepsilon/8$.

\appsubsection{DV state verification}\label{app:finite-sector}

The Stage~2 guarantee in Lem.~\ref{lem:lft} depends on the spectral gap of the finite-sector strategy and assumes that the binary measurement is implemented on the full Fock space.

\appproofheading{Lem.~\ref{lem:lft}}\label{app:tail-fidelity}

For the exact target, $P_K\ket{\psi}=\sqrt{1-q_K(\psi)}\ket{\psi_K}$ and $\Omega_K\ket{\psi_K}=\ket{\psi_K}$ by Eq.~\eqref{eq:finite-verifier}. Thus the pass probability in Eq.~\eqref{eq:lft-round} is $[1-q_K(\psi)]+q_K(\psi)=1$, proving Eq.~\eqref{eq:lift-complete}.

Now suppose $F(\rho,\psi)\leq1-\varepsilon$ and $q_K(\rho)<\varepsilon/8$. Let $R_K=P_K-\proj{\psi_K}$. Write $a=\bra{\psi_{K}}\rho\ket{\psi_{K}}$ and $r=\tr[R_{K}\rho]$, so that $a+r+q_K(\rho)=1$. If $q_K(\psi)>0$, put $\ket{\chi_{K}}=P_K^{\perp}\ket{\psi}/\sqrt{q_K(\psi)}$. Then
\begin{align}
\ket{\psi}=\sqrt{1-q_K(\psi)}\ket{\psi_{K}}
+\sqrt{q_K(\psi)}\ket{\chi_{K}}.
\label{eq:app-decomposition}
\end{align}
Applying the triangle inequality to the seminorm $\ket v\mapsto\lVert\sqrt\rho\ket v\rVert$ yields
\begin{align}\label{eq:app-triangle}
\sqrt{[1-q_K(\psi)]a}
\leq\sqrt{\bra{\psi}\rho\ket{\psi}}
+\sqrt{q_K(\psi)\bra{\chi_{K}}\rho\ket{\chi_{K}}}
\leq\sqrt{1-\varepsilon}+\sqrt{q_K(\psi)\,q_K(\rho)}.
\end{align}
The second inequality uses $\proj{\chi_K}\leq P_K^\perp$, which gives $\bra{\chi_K}\rho\ket{\chi_K}\leq q_K(\rho)$ for positive $\rho$. When $q_K(\psi)=0$, the same bound follows from $\ket{\psi_K}=\ket\psi$. With $q_K(\rho)<\varepsilon/8$ and $q_K(\psi)\leq\varepsilon/8$, the bound on $a$ gives
\begin{align}\label{eq:app-small-tail-weight}
r=1-q_K(\rho)-a
\geq 1-\frac{\varepsilon}{8}
-\frac{\left(\sqrt{1-\varepsilon}+\varepsilon/8\right)^2}{1-\varepsilon/8}
=\frac{\varepsilon-\left(\varepsilon/4\right)\left(1+\sqrt{1-\varepsilon}\right)}{1-\varepsilon/8}
\geq\frac{\varepsilon}{2}.
\end{align}
By Eq.~\eqref{eq:finite-verifier}, $\Omega_K^+\leq I-\nu_K R_K$. Taking the expectation in $\rho$ gives $\Pr_\rho(Y_i=0)\leq 1-\nu_K r\leq 1-\nu_K\varepsilon/2$, proving Eq.~\eqref{eq:conditional-gap}. Since the copies and the choices of settings are independent, $\Pr_\rho(Y=0)=[\tr[\Omega_K^+\rho]]^{N_2}$. Raising the single-round bound to the power $N_2$ and using $1-u\leq e^{-u}$ gives Eq.~\eqref{eq:lft-pass}.

\appsubsubsection{Optical implementation of the local DV measurements}\label{app:dv-implementation}

The full-space Stage~2 test in Eq.~\eqref{eq:dv-round-outcome} applies the DV decision rule within the cutoff and accepts every state supported on the tail.
We describe its local implementation and establish when Gaussian preprocessing or passive interferometry yields rejection effects diagonal in photon number.
These limits motivate an all-optical product projector strategy that uses coherent references and photon counting to probe photon number coherence.
Finally, for single-mode pure targets, we construct its rejection basis explicitly and propose a second strategy that pairs the Fock tail test with an optical SWAP test.

\begin{figure}[H]
\centering
\includegraphics[width=0.98\linewidth]{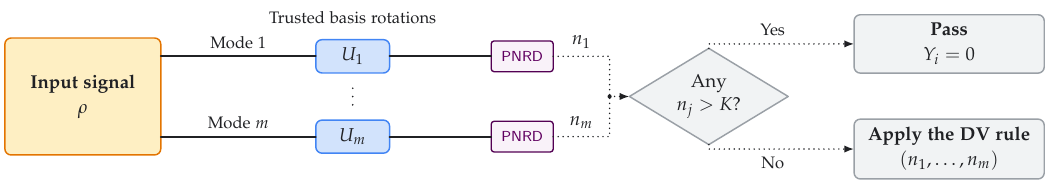}
\caption{\raggedright\textbf{Local basis measurements for a full-space Stage~2 round.}
The trusted rotations map the chosen local bases within the cutoff to photon number states and act as the identity above $K$. Dotted arrows carry the counts $n_j$ to the classical decision. If any count exceeds $K$, the round passes. Otherwise the parties apply the DV rule.}
\label{fig:stage-two-local-readout}
\end{figure}

\medskip
\textbf{Local implementation of the full-space Stage~2 test.}
For a setting implemented by local basis measurements, Fig.~\ref{fig:stage-two-local-readout} shows the order of operations. The readout assumes an exact trusted block rotation $U_j$ that maps the chosen basis of $P_{K,j}\cH_j$ to the number basis and acts as the identity above $K$. Party $j$ then uses PNRD to record $n_j$. This count checks the cutoff and, when $n_j\leq K$, identifies the DV basis outcome.

With a separate cutoff measurement, each party measures $\{P_{K,j},I_j-P_{K,j}\}$ with the corresponding projective state update. If all parties report their cutoff projector $P_{K,j}$, their joint Kraus operator is $P_K$ and the DV test can proceed. Every other outcome pattern passes immediately, so the post-measurement state on those branches is immaterial. Ordinary PNRD on the input resolves the photon number and consumes the mode. Binning the counts afterward recovers the cutoff probabilities alone.

The cutoff can also be incorporated into each party's first local instrument. At a given node of the DV protocol~\cite{Chitambar2014LOCC}, let $a$ denote the local outcome and $\mu$ index the Kraus operators $A_{a\mu}$ for that outcome. On the full input mode, use $A_{a\mu}P_{K,j}$ for its ordinary outcomes and $I_j-P_{K,j}$ for an additional local outcome indicating more than $K$ photons in that mode. This defines a complete local instrument because $\sum_{a,\mu}P_{K,j} A_{a\mu}^\dagger A_{a\mu}P_{K,j}=P_{K,j}$. After an ordinary outcome the party follows the original DV protocol. If a party obtains the additional outcome, the round passes. Before declaring a DV rejection, ask each party whose input cutoff has not yet been checked to count photons and report whether its count exceeds $K$. At this terminal step only the classical verdict matters. Thus every rejecting branch has support in $P_K\cH$, and its action there agrees with the DV setting. The rejection and pass effects on the full input space are consequently $P_K-M_{x,K}$ and $I-(P_K-M_{x,K})=P_K^\perp+M_{x,K}$, even for sources with coherence across the cutoff.

\medskip
\textbf{Diagonal rejection effects under restricted optical measurements.}
The basis readout in Fig.~\ref{fig:stage-two-local-readout} allows non-Gaussian operations such as an exact two-level rotation of $\ket{0}$ and $\ket{1}$ that fixes all higher Fock states. Krastanov \emph{et al.} give an approximate synthesis of such rotations using dispersive oscillator--qubit control~\cite[Eqs.~(9) and~(10)]{Krastanov2015Universal}. The exact cutoff-preserving action is an assumption of the readout.
Phase shifts, displacements, and squeezing are standard single-mode Gaussian gates with established optical implementations, so they are natural candidates for preprocessing before PNRD.
Can this approach realize the Stage~2 effects in Eq.~\eqref{eq:dv-round-outcome} when all settings are chosen before detection? In general, no.
In each setting, party $j$ applies a single-mode Gaussian unitary $G_j$ directly to its signal mode. Let $G=\bigotimes_j G_j$ and write $r_{\boldsymbol n}$ for the probability of rejection given the joint counts $\boldsymbol n$. The rejection effect is
\begin{align}
R_x=\sum_{\boldsymbol n} r_{\boldsymbol n}\,G^\dagger\proj{\boldsymbol n}G,
\qquad 0\leq r_{\boldsymbol n}\leq 1 .
\label{eq:gaussian-pnrd-rejection}
\end{align}
Passing every tail state requires $P_K^\perp R_x P_K^\perp=0$. Since each summand is positive, $r_{\boldsymbol n}>0$ implies $P_K^\perp G^\dagger\ket{\boldsymbol n}=0$. Thus each factor $G_j^\dagger\ket{n_j}$ of a rejecting outcome must have finite Fock support. In the Bargmann representation with complex coordinate $z$, a state has finite Fock support precisely when its wave function is a polynomial. A nonzero net displacement or squeezing of a number state introduces an exponential factor with complex quadratic and linear coefficients $a,b$, of the form $\exp(az^2+bz)$. At least one coefficient is nonzero, giving infinitely many Fock coefficients~\cite{Chabaud2019Stellar}. Thus, in a setting with a nonzero rejection effect, each $G_j$ must be a phase rotation up to an overall phase. These rotations preserve the number projectors, so the rejection effect is diagonal in the joint number basis. The same holds for any classical mixture of these settings.

A distinct measurement model uses passive interferometers, which act as multimode Gaussian unitaries, to mix each signal with local ancillas of definite photon number. It also allows classical feedforward from earlier photon counts to later optical settings. With these ancillas, conservation of total photon number at each party makes the measurement effects diagonal in the joint photon number basis independently of the tail acceptance rule (Appx.~\ref{app:perfect-completeness}).

Under the Stage~2 rule of exact tail acceptance, both nonadaptive single-mode Gaussian preprocessing followed by PNRD and passive interferometry with definite photon number ancillas and feedforward yield diagonal rejection effects.
A diagonal test that accepts $\ket{\psi_K}$ with certainty also accepts every joint photon number configuration with nonzero target amplitude. 
For a projected target $\ket{\psi_K}$ with nonzero amplitudes on two or more joint photon number configurations, the span of those configurations contains a state orthogonal to the target that also passes with certainty.
The gap within $P_K\cH$ is then zero, so \emph{a positive gap requires a measurement sensitive to coherence between the occupied configurations.}
Detector tomography has shown that mixing the signal with a coherent reference before photon detection 
can supply this sensitivity~\cite{Xu2020DetectorCoherence}. The optical construction below 
uses coherent references for this purpose and selects count records whose 
rejection effects are supported within the cutoff.

\medskip
\textbf{Product projector strategy.}
We construct a DV verification protocol for Stage~2 whose all-optical LOCC realization provides a feasible baseline under the optical assumptions below.
Each of the $m$ modes contributes $K+1$ Fock states to the cutoff sector $P_K\cH$, so the sector has dimension $(K+1)^m$, or $K+1$ for a single mode.
The sector is one-dimensional only at $K=0$, when it is spanned by the joint vacuum.
If the target has a nonzero vacuum component, the projected target at $K=0$ is the joint vacuum, so the DV test accepts the whole sector with $\Omega_0=P_0$.
For $K\geq1$, Prop.~2.1 of Ref.~\cite{Yoshida2021ProductBasis} guarantees a basis of product vectors for the orthogonal complement of $\ket{\psi_K}$ within $P_K\cH$.
The rejection vectors are the $J=(K+1)^m-1$ vectors of one such basis, denoted by $\ket{v_a}=\bigotimes_{j=1}^m\ket{v_{a,j}}$ for $a=1,\ldots,J$, with each local factor normalized.
We outline the strategy and its all-optical realization below and will give a detailed treatment for arbitrary finite-dimensional multipartite pure targets in a forthcoming paper.

In the \emph{product projector strategy}, the parties choose one rejection vector $\ket{v_a}$ uniformly from the basis in each round.
Party $j$ performs the binary measurement $\{\proj{v_{a,j}},I_j-\proj{v_{a,j}}\}$ on its mode and reports the outcome.
The parties reject exactly when each obtains the outcome associated with its local projector $\proj{v_{a,j}}$.
The joint rejection effect is therefore the product projector $\proj{v_a}=\bigotimes_{j=1}^{m}\proj{v_{a,j}}$.
This projector is fully supported in $P_K\cH$, so the strategy accepts every state on the tail.
The average pass effect restricted to $P_K\cH$ is
\begin{align}
\Omega_{K}=P_{K}-\frac{1}{J}\sum_{a=1}^{J}\proj{v_a}.
\label{eq:abstract-verifier}
\end{align}
The average rejection operator $A=J^{-1}\sum_a\proj{v_a}$ annihilates $\ket{\psi_K}$.
Let $A_\perp$ denote the restriction of $A$ to the subspace of $P_K\cH$ orthogonal to $\ket{\psi_K}$.
The rejection probability for a unit vector in this subspace is the average of its squared overlaps with the basis vectors $\ket{v_a}$.
At least one overlap is nonzero because these vectors span the subspace, so $A_\perp$ is positive definite.
The gap in Eq.~\eqref{eq:finite-verifier} is one minus the largest eigenvalue of $\Omega_K$ on this subspace, where $\Omega_K$ acts as the identity minus $A_\perp$.
Writing $\lambda_{\min}(B)$ for the smallest eigenvalue of a finite-dimensional Hermitian operator $B$, we have
\begin{align}
  \nu_K=\lambda_{\min}\left(A_\perp\right)>0.
\end{align}
This construction assumes access to the chosen local projective measurements within the cutoff.

\begin{figure}[!htp]
\centering
\vspace{2pt}
\includegraphics[width=0.98\linewidth]{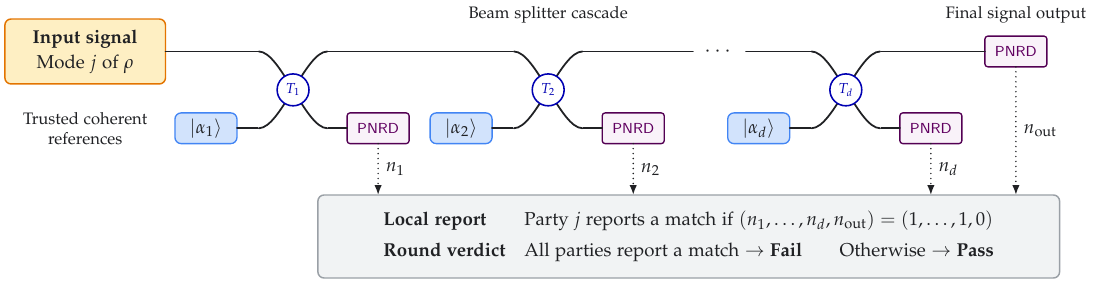}
\caption{\raggedright\textbf{Local optical circuit for the product projector strategy.}
For setting $a$, the product projector is $\proj{v_a}=\bigotimes_{j=1}^m\proj{v_{a,j}}$.
At party $j$, each splitter of power transmissivity $T_\ell$ mixes the signal with a trusted coherent reference $\ket{\alpha_\ell}$.
For the balanced choice $T_\ell=1/2$, the amplitude coefficients of Ref.~\cite{Clausen2000Overlaps} satisfy $\lvert T\rvert^2=\lvert R\rvert^2=1/2$.
The selected record $(1,\ldots,1,0)$ of tapped counts $n_\ell$ and final count $n_{\rm out}$ realizes the local effect $w_{a,j}\proj{v_{a,j}}$.
A match at every party gives the joint rejection effect $(\prod_j w_{a,j})\proj{v_a}$.}
\label{fig:product-projector-circuit}
\end{figure}

The all-optical realization of the product projector strategy uses the overlap measurement of Ref.~\cite{Clausen2000Overlaps} to implement positive multiples of the local projectors.
For party $j$ in setting $a$, let $d=d_{a,j}\leq K$ be the highest photon number with nonzero amplitude in $\vert{v_{a,j}}\rangle$. For $d\geq1$, the circuit in Fig.~\ref{fig:product-projector-circuit} mixes the signal successively with $d$ coherent reference pulses at beam splitters and measures every output by PNRD. The selected record contains one photon in each tapped output and zero in the final signal output. For example, choose identical splitters with power transmissivity $T_\ell=1/2$ ($\lvert T\rvert^2=\lvert R\rvert^2=1/2$ in the amplitude notation of Ref.~\cite{Clausen2000Overlaps}). Use the polynomial roots and reference amplitudes in Ref.~\cite[Eqs.~(17) and~(22)]{Clausen2000Overlaps} to set the pulses. The selected record then has effect $w_{a,j}\proj{v_{a,j}}$, with $0<w_{a,j}\leq1$ given by Eq.~(23) of that reference. Since the record contains $d$ photons in total, photon number conservation confines its effect to local photon numbers at most $d$. For $d=0$, the selected record is a vacuum outcome under direct PNRD.
Note that this all-optical realization assumes trusted coherent reference pulses 
with the prescribed amplitudes and phases, lossless beam splitters, and ideal PNRD at every output.

The parties reject a round when every party obtains its selected record and pass on every other outcome. With $w_a=\prod_{j=1}^m w_{a,j}>0$, the joint rejection effect is the positive multiple $w_a\proj{v_a}$ of the product projector. The average rejection operator is
\begin{align}
A_{\rm opt}=\frac{1}{J}\sum_{a=1}^{J}w_a\proj{v_a}.
\label{eq:optical-product-rejection}
\end{align}
Each optical rejection effect annihilates the projected target and vanishes on the tail, so the full target and all tail states pass with certainty.
Since all $w_a$ are positive, the inequality $A_{\rm opt}\geq(\min_a w_a)A$ ensures a positive gap at each fixed cutoff.
The optical gap $\nu_K^{\rm opt}$ is the smallest eigenvalue of $A_{\rm opt}$ on the projected target's orthogonal complement in $P_K\cH$. Since the complement has dimension $J$ and $\tr[A_{\rm opt}]=J^{-1}\sum_a w_a\leq1$, the gap satisfies $\nu_K^{\rm opt}\leq1/J$. If the target has unbounded Fock support in the measurement frame, $J$ grows as $\varepsilon$ decreases. At fixed confidence, the Stage~2 allocation in Prot.~\ref{prot:two-stage} for this optical strategy then grows at least as $J/\varepsilon$.

\medskip
\textbf{Product projector strategy for single-mode pure states.}
When $m=1$, we obtain an experimentally feasible projector strategy for arbitrary known \emph{single-mode pure states}.
For Stage~2 at $K\geq1$, use the Fock basis $\{\ket{n}\}_{n=0}^{K}$ of $P_K\cH$ to write $\ket{\psi_K}=\sum_{n=0}^{K}c_n\ket{n}$.
Choose the global phase of $\ket{\psi_K}$ so that $c_0$ is real and nonnegative.
The subspace of $P_K\cH$ orthogonal to $\ket{\psi_K}$ has dimension $J=K$.
To obtain a concrete choice of rejection vectors, we construct an orthonormal basis of this subspace directly from the coefficients $c_n$. Set
\begin{align}
  \ket{u}=\left(\ket{0}+\ket{\psi_K}\right)/\sqrt{2\left(1+c_0\right)}.
\end{align}
The reflection $P_K-2\proj{u}$ sends $\ket{0}$ to $-\ket{\psi_K}$.
For each photon number $a=1,\ldots,K$, the same reflection maps the Fock state $\ket{a}$ to the rejection vector
\begin{align}
  \ket{v_a}=\ket{a}-\frac{c_a^*}{1+c_0}\left(\ket{0}+\ket{\psi_K}\right),
  \label{eq:single-mode-rejection-basis}
\end{align}
where $c_a^*$ is the complex conjugate of $c_a$.
The squared norm of $\ket{0}+\ket{\psi_K}$ is $2(1+c_0)$, so $\ket{u}$ is a unit vector in $P_K\cH$.
The reflection is self-adjoint and satisfies $(P_K-2\proj{u})^2=P_K$, so it is unitary on this subspace.
It maps the Fock basis to $-\ket{\psi_K},\ket{v_1},\ldots,\ket{v_K}$.
Rephasing the first vector gives the complete orthonormal basis $\{\ket{\psi_K},\ket{v_1},\ldots,\ket{v_K}\}$ of $P_K\cH$.
For example, if $\ket{\psi_1}=(\ket{0}+\ket{1})/\sqrt{2}$, Eq.~\eqref{eq:single-mode-rejection-basis} gives $\ket{v_1}=(\ket{1}-\ket{0})/\sqrt{2}$.

In each round, choose $a$ uniformly and tune the coherent reference amplitudes in Fig.~\ref{fig:product-projector-circuit} for $\ket{v_a}$.
When $\ket{v_a}$ has a nonzero amplitude at a positive photon number, reject on the record of one photon in each tapped output and zero in the final signal output. This record has effect $w_a\proj{v_a}$.
For a vacuum basis vector, use direct PNRD and reject on the vacuum outcome.
All other records pass, so the exact target and states on the tail pass Stage~2 with certainty.
Eq.~\eqref{eq:optical-product-rejection} gives the optical gap $\nu_K^{\rm opt}=(\min_{1\leq a\leq K} w_a)/K\leq1/K$, which determines the Stage~2 allocation in Prot.~\ref{prot:two-stage}.

\medskip
\textbf{Optical SWAP test for single-mode pure states.}
We propose a second verification strategy for single-mode pure states by combining the Fock tail test with an optical SWAP test in Stage~2.
The product projector strategy uses coherent reference pulses to test the rejection vectors, and its gap depends on their optical weights.
The optical SWAP strategy compares each source copy with a fresh, trusted copy of the projected target and gives a gap of $1/2$ within $P_K\cH$ at every cutoff.
A threshold on the Stage~2 rejection fraction accounts for possible failures from the exact target's Fock tail.

\begin{figure}[!htp]
\centering
\includegraphics[width=0.98\linewidth]{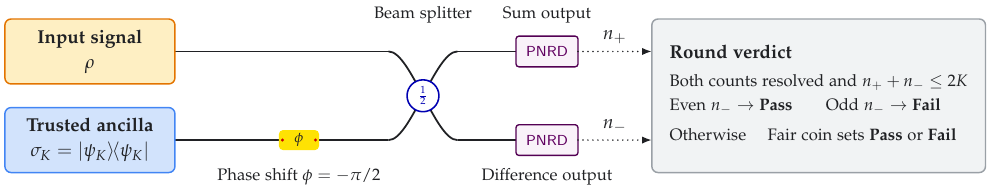}
\caption{\raggedright\textbf{Optical SWAP test for a single-mode pure state.} The unknown single-mode input $\rho$ and an independently prepared, trusted copy $\sigma_K$ of the known projected target interfere at a balanced beam splitter. In the symmetric beam splitter convention, the phase shift $\phi=-\pi/2$ gives the difference mode at the lower output up to an overall phase. The two detectors resolve the sum and difference counts $n_+$ and $n_-$ through $2K$ and flag larger counts. Dotted arrows carry the detector records to the ``Round verdict'' box.}
\label{fig:single-mode-reference-circuit}
\end{figure}

In each round, the unknown source copy $\rho$ and the known reference copy $\sigma_K=\proj{\psi_K}$ enter the setup in Fig.~\ref{fig:single-mode-reference-circuit}. The optical transformation preserves total photon number and maps the SWAP operator $V$ to photon number parity at the difference output~\cite{Jeong2014Overlap,Volkoff2022Swap}.

Let $\Pi_{\leq2K}$ project onto total photon number at most $2K$. Give a pass the signed verdict $+1$ and a fail the verdict $-1$. Assigning a fair decision above $2K$ sets the mean signed verdict there to zero. In the single-mode Fock basis, write $\rho_{ij}=\bra i\rho\ket j$ and $(\sigma_K)_{ji}=\bra j\sigma_K\ket i$. The signed verdict on the input measures $V\Pi_{\leq2K}$ and has expectation
\begin{align}
\tr[\left(\rho\otimes\sigma_K\right)V\Pi_{\leq2K}]
&=\sum_{i+j\leq2K}\rho_{ij}\left(\sigma_K\right)_{ji}
=\tr[\rho\sigma_K].
\label{eq:single-finite-counter-identity}
\end{align}
The second equality holds because every nonzero matrix element of $\sigma_K$ has $i,j\leq K$. Thus the pass effect is valid for arbitrary source energy and satisfies
\begin{align}
T_K&=\tfrac12\left(I+\sigma_K\right)\leq\tfrac12\left(P_K+\sigma_K\right)+P_K^{\perp}
\label{eq:single-swap-lift-bound},\\
{\Pr}_\rho\left(\mathrm{pass}\right)
&=\tr[\rho T_K]=\tfrac12\left(1+\tr[\rho\sigma_K]\right).
\label{eq:single-swap}
\end{align}
Choose $K$ with $q_K(\psi)\leq\tau_K\leq\varepsilon/32$. A round on the exact target fails with probability $q_K(\psi)/2\leq\tau_K/2$. The upper bound on $T_K$ has an in-sector gap of $1/2$, so Eq.~\eqref{eq:conditional-gap} gives a rejection probability of at least $\varepsilon/4$ for an $\varepsilon$-bad source with $q_K(\rho)<\varepsilon/8$.

At a common overall error budget $\delta$, allocating $\delta/2$ to false rejection in each stage suffices by the union bound. Take $N_1=\lceil96\ln(2/\delta)/\varepsilon\rceil$ for the FTT. Stage~2 passes if the failure fraction is at most $s$, with $\tau_K/2<s<\varepsilon/4$. For probabilities $u,v$, write $D(u\Vert v)=u\ln(u/v)+(1-u)\ln[(1-u)/(1-v)]$ for their binary relative entropy, with endpoint values defined by continuity. Chernoff bounds give
\begin{align}
N_2\geq\left\lceil\max\left\{
\frac{\ln\left(2/\delta\right)}{D\left(s\middle\Vert\tau_K/2\right)},
\frac{\ln\left(1/\delta\right)}{D\left(s\middle\Vert\varepsilon/4\right)}
\right\}\right\rceil.
\label{eq:single-swap-samples}
\end{align}
The first term bounds the false rejection probability in Stage~2 by $\delta/2$. If $\tau_K=0$, the first term is zero because the target always passes the SWAP round and $D(s\Vert0)=+\infty$ for $s>0$. The second term bounds acceptance of an $\varepsilon$-bad source with $q_K(\rho)<\varepsilon/8$ by $\delta$. Lem.~\ref{lem:ftt} covers the remaining sources.

\appsubsection{Two-stage verification protocol}\label{app:two-stage-protocol}\label{app:guarantee}

For a fixed i.i.d. source, let $p=\tr[\Omega_K^+\rho]$ be the single-round pass probability in Stage~2.
The binomial coefficient $\binom{N_1}{\ell}$ counts the ways to choose $\ell$ flagged copies among the $N_1$ copies tested in Stage~1.
Independent fresh copies for the two stages and independent setting choices give the exact acceptance probability
\begin{align}
{\Pr}_\rho\left(\mathrm{Good}\right)=
\left[\sum_{\ell=0}^{\left\lfloor N_1\varepsilon/16\right\rfloor}
\binom{N_1}{\ell}q_K(\rho)^\ell\left[1-q_K(\rho)\right]^{N_1-\ell}\right]p^{N_2}.
\label{eq:app-exact-two-stage}
\end{align}
Running Stage~2 only after the FTT accepts gives the same probability. The FTT measurement need not preserve coherence because Stage~2 uses fresh copies. Both tests require copies from the same source because the same $\rho$ must determine both $q_K(\rho)$ and $p$. We now bound the two error probabilities without estimating either quantity exactly.

\appproofheading{Thm.~\ref{thm:two-stage}}\label{app:complete-protocol}

For the exact target, Stage~2 passes with certainty by Lem.~\ref{lem:lft}. With $N_1$ as in Prot.~\ref{prot:two-stage}, Lem.~\ref{lem:ftt} gives a rejection probability of at most $e^{-N_1\varepsilon/96}\leq\delta_{\rm G}$. Now fix an $\varepsilon$-bad $\rho$. If $q_K(\rho)\geq\varepsilon/8$, the probability of accepting both stages is at most the FTT pass probability, which is bounded by $e^{-N_1\varepsilon/64}\leq\delta_{\rm B}$. If $q_K(\rho)<\varepsilon/8$, acceptance requires all Stage~2 rounds to pass. Lem.~\ref{lem:lft} and the protocol's choice of $N_2$ give
\begin{align}
{\Pr}_\rho\left(\mathrm{Good}\right)\leq
\left(1-\nu_K\varepsilon/2\right)^{N_2}\leq e^{-N_2\nu_K\varepsilon/2}\leq\delta_{\rm B}.
\label{eq:app-complete-low-tail}
\end{align}
The last inequality uses $N_2\geq2\ln(1/\delta_{\rm B})/(\nu_K\varepsilon)$, which also holds when a certified lower bound on the gap is used in the allocation. The two cases cover every source state without an energy assumption. Using the exact gap, $0<\nu_K\leq1$ implies $\ln(1/\delta_{\rm B})/\nu_K\geq\ln(1/\delta_{\rm B})$. The two allocations therefore sum to at most $98/\varepsilon$ times the maximum in Eq.~\eqref{eq:total-copies} before rounding. Rounding up adds less than two copies. The same budget calculation applies with a certified lower bound in place of $\nu_K$.

\appsection{Applications to Gaussian States}\label{app:gaussian-applications}

We prove the corollaries of Sec.~\ref{sec:gaussian-applications}. For entangled two-mode Gaussian targets, we specify the local measurements in the one-way DV test and calculate the acceptance probabilities in Fig.~\ref{fig:tmsv-error-acceptance}.
The corollary proofs for both Secs.~\ref{sec:gaussian-applications} and~\ref{sec:non-gaussian-applications} use the shared error budget $\delta$ and the Stage~1 sample size in Eq.~\eqref{eq:examples-ftt-samples}.

\appsubsection{Arbitrary pure multimode Gaussian states}\label{app:multimode-Gaussian-states}

\appproofheading{Cor.~\ref{cor:multimode-gaussian-verification}}\label{app:multimode-gaussian-proof}

The product unitary $L$ preserves the fidelity between the source and the target. In the transformed frame, Eq.~\eqref{eq:graph-tail} ensures $q_K(\widetilde G)\leq\varepsilon/32$ for the chosen cutoff, so the Stage~1 allocation in Eq.~\eqref{eq:examples-ftt-samples} suffices. By Eq.~\eqref{eq:graph-general-gap}, $2/\nu_K\leq2^m$. Thus $\lceil 2^m\ln(1/\delta)/\varepsilon\rceil$ copies suffice for Stage~2. Thm.~\ref{thm:two-stage} gives both acceptance guarantees. Adding the two allocations with the rounding bound gives the sufficient total $N_{\rm Gauss}$ in Eq.~\eqref{eq:multimode-gaussian-samples}.

\appsubsection{Arbitrary pure two-mode Gaussian states}\label{app:two-mode-Gaussian-states}

For the TMSV target with $r>0$ in Sec.~\ref{sec:two-mode-Gaussian-states}, the joint distribution is $(1-\lambda)\lambda^n$ on $\ket{n,n}$ and vanishes on unequal photon numbers. Since the two counts agree, the probability that both are at most $K$ is
\begin{align}
  \sum_{n=0}^K\left(1-\lambda\right)\lambda^n=1-\lambda^{K+1}.
\end{align}
Dividing each weight $(1-\lambda)\lambda^n$ for $n\leq K$ by this probability gives the Schmidt probabilities in Eq.~\eqref{eq:tmsv-truncated}.

\textbf{One-way DV verification.}
The tests in Eq.~\eqref{eq:tmsv-finite-verifier} adapt the protocol of Yu, Shang, and G\"uhne~\cite{Yu2019Optimal} to the finite Fock sector. On fresh CV copies, implement each setting by the full-space rule in Eq.~\eqref{eq:dv-round-outcome}.
With probability $w_K=p_0/(1+p_0)$, the parties use the Schmidt basis test. Both count photons in their local Fock bases, and $A$ sends its count to $B$. On the full space, a count above $K$ in either mode gives a pass. If both counts are at most $K$, party $B$ passes the copy if they match. This gives the in-sector pass effect $P_{\rm diag}$.

With probability $1-w_K$, use the Fourier test. On full-space inputs, use the coherent cutoff measurement in Appx.~\ref{app:dv-implementation} to pass above-cutoff outcomes and preserve coherence for the in-sector test. In that test, party $A$ draws $z_0,\ldots,z_K$ independently and uniformly from $\{1,\mathrm{i},-1,-\mathrm{i}\}$. Let $d=K+1$ be the local dimension, set $\omega=e^{2\pi\mathrm{i}/d}$, and write $\boldsymbol z=(z_0,\ldots,z_K)$ for the sampled phases. For $a=0,\ldots,K$, the measurement basis vectors at $A$ and the conditional pass vectors at $B$ are
\begin{align}
\ket{f_{\boldsymbol z,a}}
&=\frac{1}{\sqrt{d}}\sum_{n=0}^K z_n\omega^{an}\ket{n},\\
\ket{b_{\boldsymbol z,a}}
&=\sum_{n=0}^K\sqrt{p_n}\,z_n^*\omega^{-an}\ket{n}.
\end{align}
Party $A$ measures in the Fourier basis $\{\ket{f_{\boldsymbol z,a}}\}_{a=0}^K$ and sends $\boldsymbol z$ and the outcome $a$ to $B$. Party $B$ performs a binary test with pass projector $\proj{b_{\boldsymbol z,a}}$. On the target $\ket{\psi_K}$, $\ket{b_{\boldsymbol z,a}}$ is the normalized conditional state at $B$, so the target passes with certainty.

For fixed $\boldsymbol z$, the joint pass effect within $P_K\cH$ is the sum of $\proj{f_{\boldsymbol z,a}}\otimes\proj{b_{\boldsymbol z,a}}$ over $a$. Its matrix element between $\ket{i,j}$ and $\ket{k,l}$ has phase factor $z_i z_l z_j^*z_k^*$. Independence and the vanishing first and second moments of each phase leave only the cases $i=j$ and $k=l$, or $i=k$ and $j=l$. The Fourier phases cancel in both cases. The surviving coefficients are $\sqrt{p_i p_k}$ for $\ket{i,i}\bra{k,k}$ and $p_j$ for $\proj{i,j}$ with $i\ne j$, giving $\Omega_K^{\rm F}$ in Eq.~\eqref{eq:tmsv-finite-verifier}.

The gap of Eq.~\eqref{eq:tmsv-finite-verifier} can be read from three invariant subspaces. The target has eigenvalue one. The diagonal subspace orthogonal to the target has eigenvalue $w_K=p_0/(1+p_0)$, and each basis vector $\ket{i,j}$ with unequal photon counts has eigenvalue $(1-w_K)p_j\leq w_K$, since $p_j\leq p_0$. Therefore the largest eigenvalue on the target complement is $w_K$ and the gap is $1/(1+p_0)$. The known local Gaussian inverse transfers these statements to the original target only when it precedes both stages; it does not preserve an untransformed Fock cutoff.

\appproofheading{Cor.~\ref{cor:gaussian-verification}}\label{app:gaussian-proof}

For $r>0$, Eq.~\eqref{eq:tmsv-tail} and Eq.~\eqref{eq:tmsv-cutoff} ensure $q_K(\TMSV(r))\leq\varepsilon/32$, and Eq.~\eqref{eq:tmsv-uniform-gap} gives $\nu_K=1/(1+p_0)$. Substituting the gap bound in Prot.~\ref{prot:two-stage} and evaluating $p_0$ from Eq.~\eqref{eq:tmsv-truncated} with $\lambda=\tanh^2r$ give the sufficient total $N_{\rm TMSV}$ in Eq.~\eqref{eq:gaussian-samples} for verifying $\widetilde\rho$. Unitary invariance of fidelity in Eq.~\eqref{eq:gaussian-fidelity-reduction} transfers the guarantees of Thm.~\ref{thm:two-stage} to the original target $\ket G$. The bound $p_0\leq1$ gives the scaling in Tab.~\ref{tab:family-sample-complexity}.

\appsubsubsection{Acceptance under preparation errors}\label{app:gaussian-error-acceptance}

For Fig.~\ref{fig:tmsv-error-acceptance}, take $\ket{\psi}=\ket{\TMSV(r)}$ with $\lambda=\tanh^2r=1/2$ and set $\varepsilon=\delta=0.05$. Prot.~\ref{prot:two-stage} gives $K=9$, $N_1=5\,752$, and $N_2=180$.
For a phase shift $\phi$ on mode $A$, set $\ket{\psi_\phi}=e^{\mathrm{i}\phi\hat n_A}\ket{\psi}$ and $\rho_\phi=\proj{\psi_\phi}$. The shift changes only the phases of the Fock amplitudes, so $q_K(\rho_\phi)=q_K(\psi)=\lambda^{K+1}$. Summing the geometric series for $\braket{\psi}{\psi_\phi}$ gives
\begin{align}
F\left(\rho_\phi,\psi\right)
&=\frac{\left(1-\lambda\right)^2}{1+\lambda^2-2\lambda\cos\phi}.
\label{eq:gaussian-phase-error-fidelity}
\end{align}
We choose $\phi=\arccos(75/76)$, which gives $F(\rho_\phi,\psi)=0.95$.
Define the normalized projection $\ket{\psi_{\phi,K}}=P_K\ket{\psi_\phi}/\sqrt{1-q_K(\psi)}$. The truncated geometric sum gives
\begin{align}
\left|\braket{\psi_K}{\psi_{\phi,K}}\right|^2
&=F\left(\rho_\phi,\psi\right)
\frac{\left|1-\lambda^{K+1}e^{\mathrm{i}(K+1)\phi}\right|^2}{\left(1-\lambda^{K+1}\right)^2}.
\label{eq:gaussian-phase-error-projection}
\end{align}
Let $p(\rho)$ be the Stage~2 round pass probability in Eq.~\eqref{eq:lft-round}. The component of $\ket{\psi_{\phi,K}}$ orthogonal to $\ket{\psi_K}$ lies in the matched photon count subspace, where Eq.~\eqref{eq:tmsv-finite-verifier} assigns rejection probability $\nu_K$ to that component. Including the weight of the projection gives
\begin{align}
p\left(\rho_\phi\right)
&=1-\nu_K\left[1-q_K\left(\psi\right)\right]
        \left[1-\left|\braket{\psi_K}{\psi_{\phi,K}}\right|^2\right].
\label{eq:gaussian-phase-error-pass}
\end{align}

For high-photon contamination, let $\rho_{\rm tail}=(1-\varepsilon)\proj{\psi}+\varepsilon\proj{K+1,0}$. The Fock state $\ket{K+1,0}$ is orthogonal to $\ket{\psi}$ because its counts are unequal, so $F(\rho_{\rm tail},\psi)=1-\varepsilon$. Its contribution to the source tail gives $q_K(\rho_{\rm tail})=(1-\varepsilon)q_K(\psi)+\varepsilon$. Both components pass Stage~2 with certainty, giving $p(\rho_{\rm tail})=1$.

For either source, let $a$ be the binomial probability of at most $17$ tail flags, as prescribed by Eqs.~\eqref{eq:ftt-count} and~\eqref{eq:ftt-decision}. Let $b=p(\rho)^{N_2}$ be the probability that all Stage~2 rounds pass when Stage~1 passes. The independent batches in Appx.~\ref{app:two-stage-protocol} give $a$ for the fraction of runs that continue to Stage~2 and $ab$ for the fraction of all runs that pass both stages. For $\rho_\phi$, these are $a\simeq0.999976$ and $ab\simeq2.88\times10^{-3}$. For $\rho_{\rm tail}$, $b=1$ gives $a=ab\simeq1.62\times10^{-103}$.

\appsection{Applications to Non-Gaussian States}\label{app:non-gaussian-applications}

We prove the corollaries of Sec.~\ref{sec:non-gaussian-applications} and give the photon tail calculations used to choose the cutoffs.
We also give preparation circuits for the photon-subtracted and photon-added targets, compare verification protocols for entangled coherent states, and specify the logical measurements for NOON states.

\appsubsection{Symmetric photon subtraction from a TMSV}\label{app:photon-subtraction}

\begin{figure}[!htp]
\centering
\includegraphics[width=0.98\linewidth]{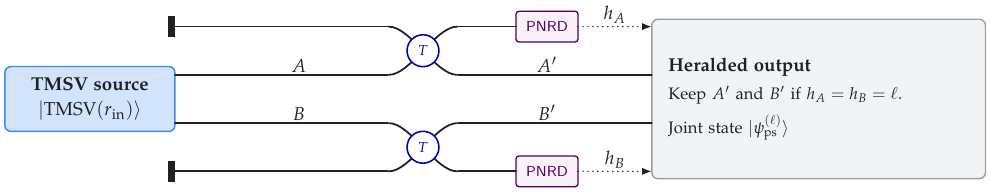}
\caption{\raggedright\textbf{Heralded preparation by symmetric photon subtraction.}
Beam splitters of intensity transmissivity $T$ tap the input modes $A$ and $B$. Black blocks mark the auxiliary input ports left in vacuum. The PNRDs record photon counts $h_A$ and $h_B$ in the reflected taps. The count condition $h_A=h_B=\ell$ heralds the target on the transmitted modes $A'$ and $B'$.}
\label{fig:photon-subtraction-circuit}
\end{figure}

Fig.~\ref{fig:photon-subtraction-circuit} shows how to prepare the state in Eq.~\eqref{eq:ps-family} from a TMSV with input squeezing $r_{\rm in}$~\cite{NavarreteBenlloch2012Enhancing}.
With ideal components, conditioning on $\ell$ photons in each tap weights the ideal subtraction amplitude of $\ket{n,n}$ by $T^n$ up to a common factor, where $T$ is the beam splitter intensity transmissivity. To obtain the target at squeezing $r$, choose $\tanh r<T<1$ and set $\tanh r_{\rm in}=\tanh r/T$.

Applying $a_A^\ell a_B^\ell$ to the TMSV at the target squeezing $r$ and reindexing the sum gives the unnormalized vector
\begin{align}
a_A^\ell a_B^\ell\ket{\TMSV\left(r\right)}
=\sqrt{1-\lambda}\,\ell!\,\lambda^{\ell/2}
\sum_{n\geq0}\binom{n+\ell}{\ell}\lambda^{n/2}\ket{n,n}.
\label{eq:app-ps-operation}
\end{align}
Its squared norm is $(1-\lambda)(\ell!)^2\lambda^\ell Z_\ell(\lambda)$. Write ${}_2F_1$ for the Gauss hypergeometric function. The generating series $Z_\ell={}_2F_1(\ell+1,\ell+1;1;\lambda)$ reduces, by the Euler transformation, to $(1-\lambda)^{-2\ell-1}{}_2F_1(-\ell,-\ell;1;\lambda)$. The latter series terminates and yields Eq.~\eqref{eq:photon-normalization}. Differentiating the convergent positive series gives the stated total mean photon number. The one-mode reduced state is diagonal in photon number. Its probability $p_n^{(\ell)}$ of $n$ photons obeys
\begin{align}
\frac{p_{n+1}^{\left(\ell\right)}}{p_n^{\left(\ell\right)}}
=\lambda\left(\frac{n+\ell+1}{n+1}\right)^2.
\label{eq:app-photon-probability-ratio}
\end{align}
For $\ell\geq1$ this ratio is not constant. A phase-invariant one-mode Gaussian state is thermal and has a geometric photon distribution, so this marginal, and hence its parent pure state, is non-Gaussian. More than one Schmidt coefficient is nonzero when $\lambda>0$, establishing entanglement.

For the general tail bound, factor $\lambda^n=(\lambda/z)^nz^n$ in Eq.~\eqref{eq:ps-tail}. On $n>K$, the first factor is at most $(\lambda/z)^{K+1}$; extending the remaining positive sum to all $n\geq0$ proves Eq.~\eqref{eq:ps-exponential-tail}. Eq.~\eqref{eq:ps-cutoff} follows by taking logarithms. It is a sufficient cutoff and need not minimize the dimension of the finite Fock sector.

For $\ell=1$, writing the tail index as $n=K+1+j$ and evaluating the zeroth, first, and second geometric moments gives
\begin{align}
q_K\left(\psi_{\rm ps}\right)=\frac{\lambda^{K+1}}{1+\lambda}
\left[\left(K+2\right)^2\left(1-\lambda\right)^2+2\left(K+2\right)\lambda\left(1-\lambda\right)+\lambda\left(1+\lambda\right)\right].
\label{eq:app-ps-closed-tail}
\end{align}
For general integer $\ell$, Eq.~\eqref{eq:photon-normalization} and the finite sum in Eq.~\eqref{eq:ps-schmidt} also give the exact tail $1-Z_\ell(\lambda)^{-1}\sum_{n=0}^K\binom{n+\ell}{\ell}^2\lambda^n$. Directed rounding or exact arithmetic is needed when a numerical value is used as a certified upper bound.

Normalizing the weights within the cutoff leaves adjacent probability ratios unchanged. For $0\leq n<K$, the probabilities in Eq.~\eqref{eq:ps-schmidt} satisfy $p_{n+1}^{(\ell,K)}\geq p_n^{(\ell,K)}$ exactly when $n+1\leq\ell\sqrt{\lambda}/(1-\sqrt{\lambda})$. A maximizing index for $p_{\max}^{(\ell,K)}$ is therefore
\begin{align}
n_* =\min\left\{K,\left\lfloor
\frac{\ell\sqrt\lambda}{1-\sqrt\lambda}\right\rfloor\right\}.
\label{eq:app-photon-mode}
\end{align}
Two neighboring probabilities can tie if the fraction is an integer. The displayed index still attains the maximum. With the mixing weight $w_K=p_{\max}^{(\ell,K)}/(1+p_{\max}^{(\ell,K)})$, the one-way effect has eigenvalue $w_K$ on the diagonal subspace orthogonal to the target and eigenvalues $(1-w_K)p_j^{(\ell,K)}\leq w_K$ on basis vectors with unequal photon counts. The target eigenvalue is one. This proves Eq.~\eqref{eq:ps-gap}, with gap one for a one-dimensional sector.

\appproofheading{Cor.~\ref{cor:photon-subtraction}}\label{app:ps-proof}

The selected cutoff bounds the exact target tail by $\varepsilon/32$. The normalized projected target has the Schmidt probabilities in Eq.~\eqref{eq:ps-schmidt}, and the one-way test has gap at least $1/(1+p_{\max}^{(\ell,K)})$. Substituting this gap bound in Prot.~\ref{prot:two-stage} gives the stated sample complexity. Thm.~\ref{thm:two-stage} gives both error guarantees. The estimate $p_{\max}^{(\ell,K)}\leq1$ gives a uniform upper bound on the sample complexity of Stage~2. These steps use no source Gaussianity or energy constraint.

\appsubsection{Symmetric photon addition to a TMSV}\label{app:photon-addition}

\begin{figure}[!htp]
\centering
\includegraphics[width=0.98\linewidth]{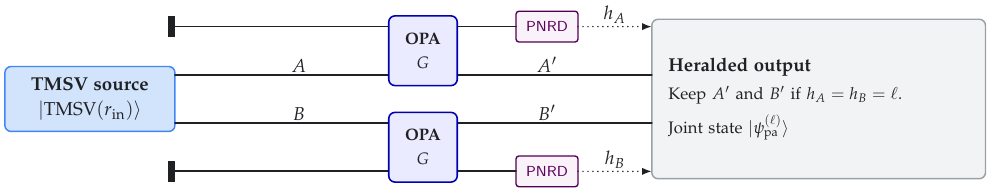}
\caption{\raggedright\textbf{Heralded preparation by symmetric photon addition.}
Each nondegenerate optical parametric amplifier (OPA) of intensity gain $G$ couples one TMSV mode to an idler. Black blocks mark the idler inputs left in vacuum. The PNRDs record the idler counts $h_A$ and $h_B$. The count condition $h_A=h_B=\ell$ heralds the target on the signal outputs $A'$ and $B'$.}
\label{fig:photon-addition-circuit}
\end{figure}

Fig.~\ref{fig:photon-addition-circuit} uses identical nondegenerate optical parametric amplifiers to prepare the target in Eq.~\eqref{eq:pa-family}~\cite{NavarreteBenlloch2012Enhancing}.
With ideal components, conditioning on $\ell$ photons in each vacuum-seeded idler weights the ideal addition amplitude of $\ket{n+\ell,n+\ell}$ by $G^{-n}$ up to a common factor, where $G$ is the amplifier intensity gain. To obtain the target at squeezing $r$, choose $1<G<1/\tanh r$ and set the input squeezing by $\tanh r_{\rm in}=G\tanh r$.

At the target squeezing $r$, the unnormalized photon-added state is
\begin{align}
\left(a_A^\dagger\right)^\ell\left(a_B^\dagger\right)^\ell\ket{\TMSV\left(r\right)}
=\sqrt{1-\lambda}\,\ell!\sum_{n\geq0}
\binom{n+\ell}{\ell}\lambda^{n/2}\ket{n+\ell,n+\ell}.
\label{eq:app-pa-operation}
\end{align}
Its squared norm is $(1-\lambda)(\ell!)^2Z_\ell(\lambda)$, which proves the normalization in Eq.~\eqref{eq:pa-family}. Define the local number shift $S_\ell=\sum_{n\geq0}\ket{n+\ell}\bra n$. Then
\begin{align}
\ket{\psi_{\rm pa}^{\left(\ell\right)}}=\left(S_\ell\otimes S_\ell\right)
\ket{\psi_{\rm ps}^{\left(\ell\right)}},\qquad
S_\ell^\dagger S_\ell=I_j.
\label{eq:app-photon-shift}
\end{align}
Here $I_j$ acts on the single mode to which $S_\ell$ is applied. For $\ell\geq1$, $S_\ell S_\ell^\dagger=I_j-\sum_{n=0}^{\ell-1}\proj n$. The isometry preserves the Schmidt probabilities and adds $2\ell$ to the total mean photon number when applied to both modes. The local marginal is invariant under phase rotations and assigns zero probability to levels $0,\ldots,\ell-1$, which excludes a thermal Gaussian marginal. This establishes non-Gaussianity. The common nontrivial Schmidt spectrum establishes entanglement for $\lambda>0$.

For $K\geq\ell$, the projector obeys $P_{K,j}S_\ell=S_\ell P_{K-\ell,j}$, giving the tail relation in Eq.~\eqref{eq:pa-tail} and the projected state in Eq.~\eqref{eq:pa-truncated}.

For $\ell=1$ and $K\geq1$, evaluating the one-photon subtraction tail from Appx.~\ref{app:photon-subtraction} at cutoff $K-1$ gives
\begin{align}
q_K\left(\psi_{\rm pa}^{\left(1\right)}\right)
&=\frac{\lambda^K}{1+\lambda}
\left[\left(K+1\right)^2\left(1-\lambda\right)^2
+2\left(K+1\right)\lambda\left(1-\lambda\right)
+\lambda\left(1+\lambda\right)\right].
\label{eq:app-pa-closed-tail}
\end{align}
The expression also gives $q_0(\psi_{\rm pa}^{(1)})=1$, since each mode contains at least one photon.

Within $P_K$, first test the local support $\operatorname{span}\{\ket\ell,\ldots,\ket K\}$ coherently. Reject its complement and apply the shifted one-way verification protocol on the accepted support. On this support, its pass effect has the same spectrum as the pass effect for photon subtraction at cutoff $J=K-\ell$. The additional sector directions have eigenvalue zero. Hence Eq.~\eqref{eq:pa-gap} follows without assuming that $S_\ell$ is an available optical operation.

\appproofheading{Cor.~\ref{cor:photon-addition}}\label{app:pa-proof}

With $K=J+\ell$, Eq.~\eqref{eq:pa-tail} bounds the target tail by $\varepsilon/32$. On $P_K\cH$, the DV verification protocol has gap at least $1/(1+p_{\max}^{(\ell,J)})$. Substituting this gap bound in Prot.~\ref{prot:two-stage} gives the stated sample complexity, and Thm.~\ref{thm:two-stage} gives both error guarantees. The photon subtraction test at cutoff $J$ has the same Schmidt probabilities and the same sufficient gap bound, so the two protocols use the same allocation of source copies at matched cutoffs. The inequality $p_{\max}^{(\ell,J)}\leq1$ makes the bound on sample complexity uniform in the squeezing and the number of added photons per mode.

\appsubsection{Multimode entangled coherent states}\label{app:entangled-coherent-states}

The projected branch overlaps in Sec.~\ref{sec:entangled-coherent-states} follow directly from $\ket\gamma=e^{-\lvert\gamma\rvert^2/2}\sum_n\gamma^n\ket n/\sqrt{n!}$. For one mode,
\begin{align}
\bra\gamma P_{K,j}\ket\zeta
=e^{-\left(\left\lvert\gamma\right\rvert^2+\left\lvert\zeta\right\rvert^2\right)/2}
\sum_{n=0}^K\frac{\left(\gamma^*\zeta\right)^n}{n!}.
\label{eq:app-ecs-projected-overlap}
\end{align}
Because $P_K$ is a product, the diagonal contributions for each coherent branch and the cross term factor over modes. The squared norm of the projected superposition gives the probability that all counts are at most $K$. Its complement is the tail probability in Eq.~\eqref{eq:ecs-tail}. Although each branch has Poisson photon statistics, their superposition also contributes an interference term.

For the family with $\boldsymbol\beta=\boldsymbol 0$, $P_K\ket{0}^{\otimes m}=\ket{0}^{\otimes m}$, so truncation preserves the vacuum contribution and its overlap with the other branch. Eq.~\eqref{eq:ecs-tail} then gives $q_K(\Psi_{\rm ECS})=(1-\prod_{j=1}^m f_K(\alpha_j))/\mathcal Z$.

\appproofheading{Cor.~\ref{cor:ecs-verification}}\label{app:ecs-proof}

Evaluating the cutoff matrix elements in Eq.~\eqref{eq:ecs-cutoff-functions} gives the tail in Eq.~\eqref{eq:ecs-tail} and the normalized projected target in Eq.~\eqref{eq:ecs-truncated}. The stage allocations in Prot.~\ref{prot:two-stage} sum to at most the stated total. Thm.~\ref{thm:two-stage} applies to every i.i.d. source state, including those with arbitrarily large energy. For the specified adaptive test, the gap bound $\nu_K\geq2^{1-m}$ gives the scaling in Tab.~\ref{tab:family-sample-complexity}.

\appsubsubsection{Comparison with the direct photon-counting protocol}\label{app:ecs-liu-comparison}
We compare total source copies at common $\varepsilon$ and $\delta$ under ideal measurements in the i.i.d. source model of Sec.~\ref{sec:state-verification-task}. Liu \emph{et al.}~\cite{Liu2021CVQSV} test the target directly using local displacements, on/off detection, and photon number parity. Let $\nu_{\rm Liu}$ denote the largest full-space verification gap obtainable by varying the probabilities of their prescribed settings.

Consider the normalized sum of the coherent branch $\ket{\boldsymbol\alpha}$ and the vacuum branch $\ket{0}^{\otimes m}$, with $\alpha_j\ne0$ on every mode. The local displacement in Sec.~\ref{sec:entangled-coherent-states} brings the two-mode example to this form. For mode $j$, write $\mathcal L_j=\operatorname{span}\{\ket{0},\ket{\alpha_j}\}$ for its coherent-state span and $s_j=\braket{0}{\alpha_j}=e^{-\lvert\alpha_j\rvert^2/2}$ for the branch overlap. Define the displaced parity $\Pi_j^{\rm disp}=D_j(\alpha_j/2)\Pi_jD_j(-\alpha_j/2)$, where $D_j(\zeta)$ displaces mode $j$ by $\zeta$ and $\Pi_j$ acts on the local Fock state $\ket n$ with eigenvalue $(-1)^n$. The operator $\Pi_j^{\rm disp}$ swaps $\ket{0}$ and $\ket{\alpha_j}$. We call its $+1$ and $-1$ eigenspaces even and odd, respectively. Both remain infinite-dimensional when restricted to $\mathcal L_j^\perp$.

In this frame, the two settings on adjacent modes $j,j+1$ in Ref.~\cite{Liu2021CVQSV} have rejection projectors
\begin{subequations}\label{eq:app-ecs-liu-pair-effects}
\begin{align}
R_{j,1}&=\left(I_j-\proj{\alpha_j}\right)\otimes\left(I_{j+1}-\proj{0}\right),\\
R_{j,2}&=\left(I_j-\proj{0}\right)\otimes\left(I_{j+1}-\proj{\alpha_{j+1}}\right),
\end{align}
\end{subequations}
with identity on all other modes. The parity setting rejects the odd eigenspace of $\Pi_1^{\rm disp}\otimes\cdots\otimes\Pi_m^{\rm disp}$. Let $w$ be the probability of choosing that setting. Each test preserves the decomposition of each local mode into $\mathcal L_j$ and the even and odd parts of $\mathcal L_j^\perp$. On each resulting block, the operators act as finite matrices on the factors in $\mathcal L_j$ and as scalars on the other factors. Thus the averaged rejection effect has finitely many distinct eigenvalues. With all weights positive, the unique common pass state established in Ref.~\cite{Liu2021CVQSV} implies a positive verification gap. For any weights, the normalized difference of the two branches is orthogonal to the target, passes every pair test, and fails the parity test. Its rejection probability is $w$.

\paragraph{Exact two-mode gap.}
Conjugation by $\Pi_1^{\rm disp}\otimes\Pi_2^{\rm disp}$ exchanges the two pair tests and leaves the parity test and target invariant. The gap is the minimum rejection probability over unit vectors orthogonal to the target, so it is concave in the setting probabilities. Averaging a distribution with the one obtained by interchanging its pair weights gives a gap at least as large. We may therefore take both pair weights equal and denote the average of their rejection projectors by $B$.

On $\mathcal L_1\otimes\mathcal L_2$, the pair rejection projectors are rank one. Their normalized vectors can be chosen as $\ket b$ and $(\Pi_1^{\rm disp}\otimes\Pi_2^{\rm disp})\ket b$, where each local factor of $\ket b$ is orthogonal to the branch tested on that mode. Each branch has expectation $s_j$ of $\Pi_j^{\rm disp}$, which has trace zero on $\mathcal L_j$. The orthogonal unit vector in $\mathcal L_j$ therefore has expectation $-s_j$. Thus $\bra b(\Pi_1^{\rm disp}\otimes\Pi_2^{\rm disp})\ket b=s_1s_2$. The even block of $B$ is the outer product of the even component of $\ket b$ with itself. Its nonzero eigenvalue is the squared norm of that component, $(1+s_1s_2)/2$.

If mode $j$ is in $\mathcal L_j^\perp$ and mode $k$ is in $\mathcal L_k$, $B$ acts on mode $k$ as $I_k-(\proj{0}+\proj{\alpha_k})/2$. The smallest eigenvalue is $(1-s_k)/2$, on the even vector proportional to $\ket{0}+\ket{\alpha_k}$. Choosing an even vector in $\mathcal L_j^\perp$ realizes this eigenvalue in the joint even sector. If both modes lie outside their spans, $B$ is the identity. Write $s_{\max}=\max\{s_1,s_2\}$. The smallest even-sector rejection eigenvalue of $B$ orthogonal to the target is then $b=(1-s_{\max})/2$. The smallest odd-sector rejection probability for the full test is $w$, attained by the branch difference. The gap after optimizing $w$ is
\begin{align}
\nu_{\rm Liu}=\max_{0\leq w\leq1}\min\left\{w,\left(1-w\right)b\right\}
=\frac{b}{1+b}.
\label{eq:app-ecs-liu-two-mode-gap}
\end{align}
For the target proportional to $\ket{\alpha,0}+\ket{0,\beta}$ with $\alpha,\beta\ne0$, the overlaps after the local displacement are $e^{-\lvert\alpha\rvert^2/2}$ and $e^{-\lvert\beta\rvert^2/2}$. Taking the reciprocal gives
\begin{align}
\frac{1}{\nu_{\rm Liu}}=1+\frac{2}{1-s_{\max}}.
\label{eq:ecs-liu-two-mode}
\end{align}

\paragraph{Multimode gap bound.}
For $\alpha_j=\gamma\ne0$ on all modes, let $s=e^{-\lvert\gamma\rvert^2/2}$ denote the common overlap $s_j$. Choose a mode $j$ and form a test state by taking an even unit vector from $\mathcal L_j^\perp$ in mode $j$ and the normalized sum of the coherent and vacuum branches on all other modes. This state is orthogonal to the target and passes both the parity setting and all pair tests that leave mode $j$ untouched. For any pair test involving mode $j$, the local rejection projector on mode $j$ acts as the identity. The other mode is tested against one of its two local branch states. Its rejection probability is
\begin{align}
b_m(s)=\frac{1-s^2}{2\left(1+s^{m-1}\right)}.
\label{eq:app-ecs-liu-pair-rejection}
\end{align}
Indeed, the branch matched by the projector contributes zero rejection, while the other contributes $1-s^2$. The squared norm of the sum of the two branches on the remaining $m-1$ modes is $2(1+s^{m-1})$.

Every adjacent pair consists of one odd-indexed and one even-indexed mode, so the total weight of all pair tests involving odd-indexed modes is $1-w$, counting each test once. There are $\lceil m/2\rceil$ odd-indexed modes. The test state for at least one such mode is therefore rejected with probability at most $(1-w)b_m(s)/\lceil m/2\rceil$. Combining this with the rejection probability $w$ of the branch difference gives
\begin{align}
\nu_{\rm Liu}\leq
\max_{0\leq w\leq1}\min\left\{w,\frac{\left(1-w\right)b_m(s)}{\left\lceil m/2\right\rceil}\right\}
=\frac{b_m(s)}{\left\lceil m/2\right\rceil+b_m(s)}.
\label{eq:app-ecs-liu-multimode-gap}
\end{align}
Taking the reciprocal gives
\begin{align}
\frac{1}{\nu_{\rm Liu}}\geq
1+\frac{2\left\lceil m/2\right\rceil\left(1+s^{m-1}\right)}{1-s^2}.
\label{eq:ecs-liu-multimode}
\end{align}
The small gap at weak amplitudes arises from errors outside one mode's coherent-state span that pass the parity setting. A pair test involving that mode rejects only when a neighboring detector also clicks.

\paragraph{Total copy budgets.}
Every setting of the direct protocol accepts the target with certainty. An $\varepsilon$-bad source passes a single round with probability at most $1-\nu_{\rm Liu}\varepsilon$. This bound is attained by mixing the exact target with a least-rejected orthogonal state of weight $\varepsilon$. For independently sampled settings and acceptance only when every copy passes, the worst-case copy requirement is therefore
\begin{align}
N_{\rm Liu}=\left\lceil\frac{\ln\left(1/\delta\right)}
{-\ln\left(1-\nu_{\rm Liu}\varepsilon\right)}\right\rceil.
\label{eq:ecs-liu-samples}
\end{align}
This comparison covers all $\varepsilon$-bad i.i.d. sources. The numerical copy estimates in Ref.~\cite{Liu2021CVQSV} cover selected photon loss and displacement errors.

For the two-stage protocol, we choose the adaptive DV measurements of Li and Zhu~\cite{Li2026Universal}. Their gap satisfies $\nu_K\geq2^{1-m}$, so Eq.~\eqref{eq:ecs-samples} gives the sufficient budget
\begin{align}
N_{\rm ECS}\leq
\left\lceil\frac{96+2^m}{\varepsilon}\ln\left(1/\delta\right)\right\rceil+1.
\label{eq:app-ecs-uniform-copies}
\end{align}
For $m=2$, Fig.~\ref{fig:ecs-copy-comparison} compares the sufficient allocation in Eq.~\eqref{eq:app-ecs-uniform-copies} with the exact optimized worst-case requirement from Eqs.~\eqref{eq:ecs-liu-samples} and~\eqref{eq:ecs-liu-two-mode}. At $\varepsilon=\delta=0.01$, the sufficient two-stage budget is $46\,053$ copies. The plotted direct-test requirement crosses this value near common amplitude $0.202$. At $\alpha=\beta=0.1$, the direct test requires $185\,126$ copies in the worst case. At $\alpha=\beta=0.515$, it requires $7\,875$ copies, below the $44\,210$ copies allocated to Stage~1 by Eq.~\eqref{eq:examples-ftt-samples}.

At fixed $\delta$ and small $\varepsilon$, the coefficient of $\ln(1/\delta)/\varepsilon$ for the direct test approaches $1/\nu_{\rm Liu}$. The sufficient two-stage coefficient for $m=2$ is $100$, which is smaller when $s_{\max}>97/99$ or, equivalently, when the smaller of $\lvert\alpha\rvert$ and $\lvert\beta\rvert$ is below approximately $0.202$. For the GHZ-like family at fixed $m$, the lower bound in Eq.~\eqref{eq:ecs-liu-multimode} grows inversely with $\lvert\gamma\rvert^2$ at weak amplitudes, whereas the bound in Eq.~\eqref{eq:app-ecs-uniform-copies} is independent of $\gamma$. For example, at $m=3$, $\gamma=0.1$, and $\varepsilon=\delta=0.01$, $47\,895$ copies suffice for both stages, while the direct test requires at least $368\,875$.

\paragraph{Target phase and finite photon number resolution.}
For real $\alpha>0$, the midpoint displacements of Ref.~\cite{Liu2021CVQSV} map the two branches of the target proportional to $\ket{\alpha,0}+i\ket{0,\alpha}$ to $\ket{\alpha/2,-\alpha/2}$ and $\ket{-\alpha/2,\alpha/2}$. Joint parity exchanges these displaced branches. Their sum and difference both have nonzero coefficients because of the relative phase $\pi/2$, so the exact target gives both parity outcomes. The projected target in Eq.~\eqref{eq:ecs-truncated} incorporates this phase in the two-stage protocol.

In the basis readout of Appx.~\ref{app:dv-implementation}, the trusted rotations act as the identity above $K$. A detector can therefore resolve $0,\ldots,K$ and report one outcome for all counts above $K$. With reliable detection of that outcome, Stage~1 records it as a tail flag and Stage~2 counts the round as a pass. The sample complexity accounts for both types of records for any source in the i.i.d. model. For parity measurements with finite detector resolution, Ref.~\cite{Liu2021CVQSV} discards saturated records and computes the probability of a usable result.

\appsubsection{Balanced multimode NOON states}\label{app:noon}

Each Fock vector in Eq.~\eqref{eq:noon-state} has largest local count $s$. Thus the target tail is one for $K<s$ and zero at $K=s$. Let $\ket{0}_{\rm L},\ket{1}_{\rm L}$ denote the logical basis states of each qubit, encoded as $\ket0,\ket s$, respectively. This local isometry maps the qubit $W$ state to the target. Within $P_K\cH$, testing membership in each local qubit subspace extends the qubit pass effect by zero on the orthogonal complement. Hence its second-largest eigenvalue is unchanged.

Write $X_{\rm L},Y_{\rm L},Z_{\rm L}$ for the Pauli operators in this logical basis. For $m\geq3$, choose a pair of modes uniformly and measure every other logical qubit in the $Z_{\rm L}$ basis~\cite{Liu2019Dicke}. If there are no excitations among these outcomes, pass when the chosen pair has equal $X_{\rm L}$ outcomes. If there is one excitation, pass when both qubits in the pair give logical zero in $Z_{\rm L}$. Reject all other records. Liu \emph{et al.} give the $m\geq3$ gaps in Eq.~\eqref{eq:noon-gap}. For $m=2$ the state is $(\ket0_{\rm L}\otimes\ket1_{\rm L}+\ket1_{\rm L}\otimes\ket0_{\rm L})/\sqrt2$. Choose $X_{\rm L}$, $Y_{\rm L}$, or $Z_{\rm L}$ uniformly and measure both qubits in that basis. Pass if the product of the two outcomes is $+1$ for $X_{\rm L}$ or $Y_{\rm L}$, or $-1$ for $Z_{\rm L}$. Each of the other three Bell states passes exactly one setting, giving gap $2/3$.

\appproofheading{Cor.~\ref{cor:noon}}\label{app:noon-proof}

The target has zero tail at $K=s$ and always passes the logical test, so the protocol has perfect completeness for this family. Applying Thm.~\ref{thm:two-stage} to the Stage~2 tests with the gaps in Eq.~\eqref{eq:noon-gap} gives the false acceptance bound at the stated sample size. The bound $\nu_K\geq1/m$ gives the scaling in Tab.~\ref{tab:family-sample-complexity}.

\appsubsection{Finite-energy GKP Bell pairs}\label{app:gkp-bell}

Mehler's identity gives the position kernel of $e^{-2t\hat n}$. At the position coordinates $q=2\sqrt\pi\,x$ and $q'=2\sqrt\pi\,y$, this kernel is $C_t e^{-\pi H_t(x,y)}$. Summing over both comb lattices gives Eq.~\eqref{eq:gkp-bell-overlaps}. The quadratic form $H_t$ is positive definite for $t>0$, so these sums converge. Expanding the squared norm of the filtered Bell sum gives the product of the two single-mode overlaps for each pair of logical labels. The overlaps are real and $G_{01}(t)=G_{10}(t)$, which gives the cross term in Eq.~\eqref{eq:gkp-bell-normalization}.

Write $a_n^{(\mu)}=\braket{n}{\mu}_{\rm id}$ for the real Fock coefficients of the ideal combs in Sec.~\ref{sec:gkp-bell}. The same normalization in the Fock basis is
\begin{align}
b_{nk}&=\sum_{\mu=0}^1 a_n^{\left(\mu\right)}a_k^{\left(\mu\right)},\\
Z_{\rm B}\left(t\right)&=\sum_{n,k\geq0}e^{-2t\left(n+k\right)}
\left\lvert b_{nk}\right\rvert^2.
\label{eq:app-gkp-bell-fock-normalization}
\end{align}
The vacuum coefficient is positive. In the tail sum $\max\{n,k\}>K$ implies $n+k\geq K+1$. Bounding one factor $e^{-t(n+k)}$ by $e^{-t(K+1)}$ and summing the other over all $n,k$ gives Eq.~\eqref{eq:gkp-bell-tail}. In particular the least admissible cutoff can be found by summing the normalized probabilities over $0\leq n,k\leq K$, with a sufficient upper limit supplied by that inequality.

\appproofheading{Cor.~\ref{cor:gkp-bell}}\label{app:gkp-bell-proof}

On the cutoff square, the coefficient matrix $e^{-t(n+k)}b_{nk}$ is a sum of two outer products. Its singular values, after normalization, give the Schmidt amplitudes even when the two local filtered combs overlap. Applying the one-way test of Refs.~\cite{Yu2019Optimal,Li2019Efficient} in the resulting Schmidt bases gives $2/\nu_K\leq4$. Thm.~\ref{thm:two-stage} with the common Stage~1 allocation gives the stated sample complexity and both error bounds.

\appsubsection{Finitely squeezed three-mode CV hypergraph states}\label{app:cv-hypergraph}

Let $\ket{\mathrm{in}}=\ket{0;r}_p^{\otimes3}$ be the input and write $U=e^{i\gamma\hat q_1\hat q_2\hat q_3}$. For distinct modes $j,k,\ell$, the position operators commute with $U$, while $U^\dagger\hat p_j U=\hat p_j+\gamma\hat q_k\hat q_\ell$. The cross term vanishes in the input product state, giving $\langle U^\dagger\hat p_j^2 U-\hat p_j^2\rangle_{\rm in}=\gamma^2\langle\hat q_k^2\rangle_{\rm in}\langle\hat q_\ell^2\rangle_{\rm in}=\gamma^2e^{4r}/4$. Since $\hat n_j=(\hat q_j^2+\hat p_j^2-1)/2$, summing over the three modes gives Eq.~\eqref{eq:cv-hypergraph-energy}. The same transformation gives the output moment $\langle\hat q_1\hat q_2\hat p_3\rangle_{H_{\gamma,r}}=\gamma e^{4r}/4$.

\appproofheading{Cor.~\ref{cor:cv-hypergraph}}\label{app:cv-hypergraph-proof}

Whenever one count exceeds $K$, the total count is at least $K+1$. Markov's inequality for the target total count gives Eq.~\eqref{eq:cv-hypergraph-tail}. Within the cutoff, the projected target is pure and tripartite. The LOCC construction of Ref.~\cite{Li2026Universal} gives $\nu_K\geq2^{1-3}=1/4$, hence $2/\nu_K\leq8$. Substitution in Prot.~\ref{prot:two-stage} and Thm.~\ref{thm:two-stage} gives the stated sample complexity and both acceptance bounds.

\appsubsection{Single-mode non-Gaussian resources}\label{app:single-mode-practical}

We prove Cor.~\ref{cor:single-mode} and give the Fock coefficients and photon tail calculations for the cat, cubic phase, and finite-energy GKP states in Sec.~\ref{sec:single-mode-pure-states}.

\appproofheading{Cor.~\ref{cor:single-mode}}\label{app:single-mode-proof}

Eq.~\eqref{eq:target-markov} ensures that each target with finite mean photon number admits a finite cutoff satisfying Eq.~\eqref{eq:design}.
For a single mode, take $M_{x,K}=\proj{\psi_K}$ in Eq.~\eqref{eq:dv-round-outcome}. With access to trusted basis rotations, choose a unitary $U_K$ that sends $\ket{\psi_K}$ to $\ket0$ within $P_K\cH$ and acts as the identity above $K$. A sequence of two-level rotations on $\operatorname{span}\{\ket{n-1},\ket n\}$ specifies $U_K$ by zeroing the amplitude of $\ket n$ in order from $n=K$ down to $n=1$. After applying $U_K$, record the photon count $n$ with PNRD. Pass for $n=0$ or $n>K$ and reject for $1\leq n\leq K$. The resulting pass effect is $U_K^\dagger(\proj{0}+P_K^\perp)U_K=\proj{\psi_K}+P_K^\perp$.

Its restriction to $P_K\cH$ is the target projector with unit gap, so Thm.~\ref{thm:two-stage} gives the stated guarantees and sample complexity.

\appsubsubsection{Cat and cubic phase coefficients}\label{app:single-mode-coefficients}

Expanding the coherent branches of Eq.~\eqref{eq:single-cat} in the Fock basis gives
\begin{align}
c_n^\pm=\frac{e^{-\mu/2}\alpha^n\left[1\pm\left(-1\right)^n\right]}
{\sqrt{2\left(1\pm e^{-2\mu}\right)n!}}.
\label{eq:single-cat-coefficients}
\end{align}
Summing the squared moduli above $K$ gives Eq.~\eqref{eq:single-cat-tail}. The corresponding mean photon numbers are $E_+=\mu\tanh\mu$ and $E_-=\mu\coth\mu$.

For the cubic phase state, the Fock coefficients are its overlaps with the real harmonic oscillator wave functions $\phi_n$. Write $H_n$ for the physicists' Hermite polynomial of degree $n$. Then
\begin{align}
c_n=\int_{\mathbb R}\phi_n\left(q\right)\psi_{\gamma,\sigma}\left(q\right)\,dq,\qquad
\phi_n\left(q\right)=\frac{H_n\left(q\right)e^{-q^2/2}}{\pi^{1/4}\sqrt{2^nn!}}.
\label{eq:single-cubic-fock}
\end{align}
The same $\phi_n$ gives the comb coefficients below. To compute the moments in Eq.~\eqref{eq:single-moment-cutoff}, put $u(q)=\partial_q\ln\psi_{\gamma,\sigma}(q)=-q/(2\sigma^2)+3i\gamma q^2$. Define $A_j$ by $\hat n^j\psi_{\gamma,\sigma}=A_j\psi_{\gamma,\sigma}$. Primes denote differentiation with respect to $q$. Starting at $A_0=1$, the representation $\hat n=(q^2-1-\partial_q^2)/2$ yields
\begin{align}
A_{j+1}=\tfrac12\left[\left(q^2-1-u'-u^2\right)A_j-2uA_j'-A_j''\right].
\label{eq:single-cubic-moments}
\end{align}
By induction in Eq.~\eqref{eq:single-cubic-moments}, each $A_j$ is a polynomial, so $\hat n^j\psi_{\gamma,\sigma}=A_j\psi_{\gamma,\sigma}$ has finite norm. Let $Q$ have the Gaussian distribution $\mathcal N(0,\sigma^2)$ of mean zero and variance $\sigma^2$, and let $\mathbb E$ denote expectation under this distribution. Its moments of every positive integer order are finite, so $M_j=\mathbb E[A_j(Q)]$ can be evaluated using Gaussian moments. In particular, the mean photon number is
\begin{align}
E_{\gamma,\sigma}
=\frac{\sigma^2+\left(4\sigma^2\right)^{-1}+27\gamma^2\sigma^4-1}{2}.
\label{eq:single-cubic-energy}
\end{align}

\appsubsubsection{GKP coefficients and normalization}\label{app:gkp-normalization}

For the ideal square-lattice comb defined in Sec.~\ref{sec:gkp-bell}, whose filtered form appears in Eq.~\eqref{eq:single-gkp}, define
\begin{subequations}\label{eq:single-gkp-coefficients}
\begin{align}
a_n^{\left(\mu_L\right)}&=\sum_{s\in\mathbb Z}\phi_n\left(\left(2s+\mu_L\right)\sqrt\pi\right),\\
Z_{\mu_L}\left(t\right)&=\sum_{n\geq0}e^{-2tn}\left\lvert a_n^{\left(\mu_L\right)}\right\rvert^2.
\end{align}
\end{subequations}
Then $c_n=e^{-tn}a_n^{(\mu_L)}/\sqrt{Z_{\mu_L}(t)}$ are the normalized Fock coefficients. For a known logical superposition, form the corresponding linear combination of the $a_n^{(\mu_L)}$ and recompute its normalization.

Setting $\mu=\nu=\mu_L$ in Eq.~\eqref{eq:gkp-bell-overlaps} gives the positive lattice sum for the single-mode normalization
\begin{subequations}\label{eq:single-gkp-mehler}
\begin{align}
Z_{\mu_L}\left(t\right)&=C_t\sum_{s,u\in\mathbb Z}e^{-\pi H_t^{\left(\mu_L\right)}\left(s,u\right)},\\
H_t^{\left(\mu_L\right)}\left(s,u\right)&=\tanh t\,\left(s+u+\mu_L\right)^2+\coth t\,\left(s-u\right)^2.
\end{align}
\end{subequations}
Gaussian lattice remainders bound the error when this sum is truncated. For $0<q<1$, write $\vartheta_2(q)=\sum_{n\in\mathbb Z}q^{(n+1/2)^2}$ and $\vartheta_3(q)=\sum_{n\in\mathbb Z}q^{n^2}$ for the Jacobi theta constants. Set $q_a=e^{-4\pi\tanh t}$ and $q_b=e^{-4\pi\coth t}$. Changing variables to $s+u$ and $s-u$, which have the same parity, yields
\begin{subequations}\label{eq:single-gkp-theta}
\begin{align}
Z_0\left(t\right)&=C_t\left[\vartheta_3\left(q_a\right)\vartheta_3\left(q_b\right)+\vartheta_2\left(q_a\right)\vartheta_2\left(q_b\right)\right],\\
Z_1\left(t\right)&=C_t\left[\vartheta_2\left(q_a\right)\vartheta_3\left(q_b\right)+\vartheta_3\left(q_a\right)\vartheta_2\left(q_b\right)\right].
\end{align}
\end{subequations}
The mean photon number is $E_{\mu_L,t}=-\tfrac12\partial_t\ln Z_{\mu_L}(t)<\infty$. A derivative-free upper bound is
\begin{align}
E_{\mu_L,t}\leq\frac{Z_{\mu_L}\left(t/2\right)}{e\,t\,Z_{\mu_L}\left(t\right)},
\label{eq:single-gkp-energy-bound}
\end{align}
because $n e^{-tn}\leq1/(et)$. To prove Eq.~\eqref{eq:single-gkp-tail}, factor $e^{-2tn}=e^{-2hn}e^{-2(t-h)n}$ and bound the first factor in the tail sum by $e^{-2h(K+1)}$. The same argument works for a superposition of combs after substituting its normalization. Taking $h=t/2$ gives the sufficient cutoff
\begin{align}
K+1\geq\frac{\ln\left(32/\varepsilon\right)+
\ln\left[Z_{\mu_L}\left(t/2\right)/Z_{\mu_L}\left(t\right)\right]}{t}.
\label{eq:single-gkp-cutoff}
\end{align}

\end{document}